# Passive scintillometer: Theory and proof of concept.

**MIKHAIL CHARNOTSKII**

*Erie, CO, USA*
*\*Mikhail.Charnotskii@gmail.com*

**Abstract:** We propose design of a low-cost passive scintillometer that measures the strength of optical turbulence by analyzing scintillation in the image of a straight edge between the two areas of uniform, but distinct brightness. Two theoretical approaches to scintillations description are developed that are based on the phase screen and rigorous path integral propagation imaging models. Asymptotic analysis of both models leads to four distinctive imaging situations. We propose two uniform approximations that cover the most promising conclusion for the passive scintillometer applications. Both the strength of scintillation and the width of the area where scintillations exist can be used to estimate the turbulence strength. We present the results of the proof-of-concept experiment where images of the specifically made target were taken by a consumer grade camera equipped with a telephoto lens. We describe the image processing that separates the target characteristics from the turbulence scintillations. The spatial profiles of the image variances at three spectral bands and for several apertures were calculated. As was expected, scintillations are concentrated around the sharp edge and are absent at the uniformly bright parts of the target. Observed edge scintillation variances are within the theoretical limits, and provide reasonable estimates for the turbulence structure constant across the three spectral bands. However, the width of the scintillation area and dependence on the aperture are inconsistent with the thin lens imaging theory for larger apertures. We attribute this to the Additional bench tests on bar targets revealed that the camera and lens resolution are well below the nominal thin lens diffraction limit, and a different optical setup should be used in order to exploit the spatial distribution and aperture dependence of fluctuations.

## 1. Introduction

The notion of estimating the turbulence parameters from the conventional images of incoherent objects is apparent. After all, most of us are familiar with the optical effects of atmospheric turbulence from the wander and smear of some small bright details of everyday scenes we see.

Turbulence induced fluctuations in the images of extended complex objects have a very intricate dependence on the details of the object. As was shown in [1], even the simple variance of turbulent fluctuations carries a nonlocal quadratic functional dependence on the object brightness distribution. This makes it difficult to separate the pure turbulence features from the object itself in order estimate the turbulence parameters on the imaging path.

Earlier work, [2], proposed to use covariance of the turbulence OTF components at different spatial frequencies for estimation of the turbulence coherence radius $r_0$. The technique is based on the simple, phase screen, isoplanatic imaging model. Fourier contrast method, [3], also uses isoplanatic imaging model to "cancel" the object information, and estimates the turbulence strength from the variance of turbulent OTF. . ,

In [4] image fluctuations caused by translational shifts between image frames are used estimate the turbulent strength. Authors use the phase screen-based, and therefore isoplanatic [5], theory to support their technique, but in our opinion isoplanacity is not an essential constraint. Even the authors' naïve fluctuation model reflects the complex dependence of fluctuations on the local contrast: the large fluctuations areas associated are with the high-contrast areas of the scene. Only these areas are used for the image variances estimates, but formal procedure for identifying these areas are put across. Furthermore, alternative sources of image fluctuations are not accounted for.

In [6] it was proposed to estimate the turbulence strength based on the variances of the random shifts of certain patches of the image. Variations of the patch sizes and locations provide some additional regarding the turbulence distribution along propagation path. Isoplanacity is not essential to this technique, but it still does not account for the scene details by replacing the sources inside the patch by Gaussian distribution and providing no information regarding the choice of patches locations on the extended scene.

Recent work [7] proposes estimation of turbulence strength from the first and second statistical moments of the turbulent Point Spread Function (PSF). PSF samples are extracted from individual frames by blind deconvolution. Image isoplanacity is essential for his technique. In [7] it also relies on some questionable theoretical results regarding the PSF statistics. This weakness can be mitigated by employing available rigorous theoretical results of [8]. However, multi-frame blind deconvolution is computationally extensive requiring minimization of certain function of thousands of arguments for each frame, which takes hours by authors' own estimates.

Here we suggest the use of the simplest possible objects for the image-based turbulence characterization. We use the target with a straight edge separating the areas of distinct, but uniform brightness, "black" and 'white," for this purpose. This target has no spatial scales associated with it, and allows for a straightforward separation of the object features from the turbulence effects and provides for a high contrast area where turbulent fluctuations are prominent. While we used a specifically build target for our measurements described in Section 5, we also have a positive experience with the targets of opportunity, such as store signs [1].

The concept of using scintillations in the image of a sharp straight edge for turbulence characterization was proposed in [9], and path integral technique formulation for this imaging problem was presented together with some statistical parameters of the incident field from such target. Here we expand the analysis of [9] to calculate the mean image and variance of image scintillations, and present results of field measurements supporting our theoretical results.

In Section 2 we formulate the imaging problem on terms of the Green's function of random medium between the target and imaging aperture. Based on so-called "third constraint" on the turbulent PSF [5] we introduce the normalized image that is insensitive to the target brightness and contrast, and is completely defined by the turbulence and imaging aperture. In Section 3 we use the phase screen (PS) imaging model to calculate the mean image and image variance and identify two asymptotic cases: perturbation theory (PT) and coherence channels (CC). We also introduce the uniform approximation that merges these asymptotes.

PS model plays an auxiliary role in this investigation, as theoretical analysis of the image variance for the extended turbulent medium is presented in Section 4 based on the path-integral formulation for the optical field in random medium [10]. We present complete asymptotic analysis of the image variance. For the Kolmogorov turbulence we derive the set of simple working formulas that can be used for the estimation of the turbulence structure constant, $C_n^2$.

In Section 5 we report the data processing procedure and results of the field test performed using inexpensive, consumer grade camera and lens. Sections 6 summarizes the theoretical and experimental efforts, and Section 7 presents conclusions.

Five appendices in Sections 8 – 12 explain technical details of calculations used for derivation of principal theoretical results.

## 2. General discussion

In paraxial approximation wave propagation in inhomogeneous medium along *z*-axis from plane $z = 0$ can be presented as

$$u(\mathbf{r},z) = \frac{k}{2\pi i |z|} \iint d^2 r_0 u(\mathbf{r}_0, 0) \exp\left[\frac{ik}{2|z|}(\mathbf{r} - \mathbf{r}_0)^2\right] g(\mathbf{r}_0, 0; \mathbf{r}, z). \tag{1}$$

Here $u(\mathbf{r}, z)$ is the optical field at the point $\mathbf{r} = (x, y)$ in the plane $z$, and factor $g(\mathbf{r}_0, 0; \mathbf{r}, z)$ accounts for distortions caused by inhomogeneity.

It is straightforward to show, e. g. [11], that irradiance distribution in the image plane of an incoherent object can be presented as

$$I(\mathbf{R}_I) = \left(\frac{k}{2\pi L}\right)^4 \iint d^2 R_O O(\mathbf{R}_O) \iint d^2 R_A \iint d^2 r_A A\left(\mathbf{R}_A + \frac{\mathbf{r}_A}{2}\right) A\left(\mathbf{R}_A - \frac{\mathbf{r}_A}{2}\right)$$
$$\exp\left[\frac{ik}{L}(\mathbf{R}_I - \mathbf{R}_O)\cdot \mathbf{r}_A\right] g\left(\mathbf{R}_O, L; \mathbf{R}_A + \frac{\mathbf{r}_A}{2}, 0\right) g^*\left(\mathbf{R}_O, L; \mathbf{R}_A - \frac{\mathbf{r}_A}{2}, 0\right). \tag{2}$$

Here, as shown in Fig. 1, object with brightness $O(\mathbf{R}_O)$ is located in the plane $z = L$, and aperture with amplitude transmission function $A(\mathbf{R}_A)$ is, in the plane $z = 0$. Image is formed in the plane $z = -l$ by a thin lens in the aperture plane, and the object plane scaling is used for the image.

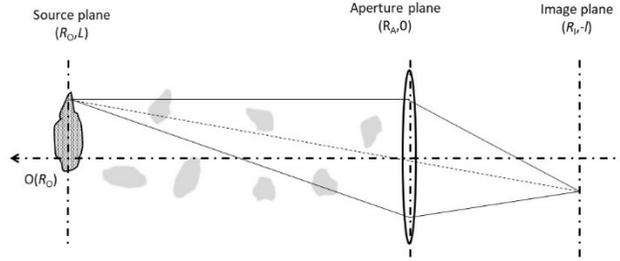

Fig. 1. Geometry of the imaging system.

A unique property of the imaging through refractive turbulence is the so-called "third PSF constraint" [5], first described in [12], or orthogonality of Green's functions, [13]. It can be formally represented as

$$\iint d^2 R_O \exp\left[\frac{ik}{L}(\mathbf{R}_I - \mathbf{R}_O)\cdot \mathbf{r}_A\right] g\left(\mathbf{R}_O, L; \mathbf{R}_A + \frac{\mathbf{r}_A}{2}, 0\right) g^*\left(\mathbf{R}_O, L; \mathbf{R}_A - \frac{\mathbf{r}_A}{2}, 0\right)$$
$$= \frac{4\pi^2 L^2}{k^2} \delta(\mathbf{r}_A), \tag{3}$$

and infers that the image of a uniform incoherent source, $O(\mathbf{R}_O) = const$, is not affected by turbulence and does not fluctuate. As a result, in the case of a "step" object

$$O(\mathbf{R}_O) = O(X_O, Y_O) = B + \frac{C}{2} \text{sign}(X_O), \tag{4}$$

the constant background, $B$, of the object results in the uniform, non-fluctuating background in the image

$$I_B = B\left(\frac{k}{2\pi L}\right)^2 \Sigma_A, \quad \Sigma_A = \iint d^2 R_A A^2(\mathbf{R}_A). \tag{5}$$

After subtracting the background, the variable component of the image is

$$\tilde{I}(X_I,Y_I) = \frac{C}{2}\left(\frac{k}{2\pi L}\right)^4 \iint dX_O dY_O \, sign(X_O) \iint d^2 R_A \iint d^2 r_A A\left(\mathbf{R}_A + \frac{\mathbf{r}_A}{2}\right) A\left(\mathbf{R}_A - \frac{\mathbf{r}_A}{2}\right)$$
$$\exp\left[\frac{ik}{L}(\mathbf{R}_I - \mathbf{R}_O)\cdot \mathbf{r}_A\right] g\left(\mathbf{R}_O, L; \mathbf{R}_A + \frac{\mathbf{r}_A}{2}, 0\right) g^*\left(\mathbf{R}_O, L; \mathbf{R}_A - \frac{\mathbf{r}_A}{2}, 0\right), \quad (6)$$

and it is expected that far away from the step image again behaves as in the case of the uniform object and does not fluctuate, namely

$$\tilde{I}(X_I \to \pm\infty, Y_I) = \pm\frac{C}{2}\left(\frac{k}{2\pi L}\right)^2 \Sigma_A. \quad (7)$$

After normalization, by the contrast, the normalized instantaneous image

$$i(X_I,Y_I) \equiv \frac{\tilde{I}(X_I,Y_I)}{\tilde{I}(\infty,Y_I)} = \frac{k^2}{4\pi^2 L^2 \Sigma_A} \iint dX_O dY_O \, sign(X_O) \iint d^2 R_A \iint d^2 r_A A\left(\mathbf{R}_A + \frac{\mathbf{r}_A}{2}\right)$$
$$\times A\left(\mathbf{R}_A - \frac{\mathbf{r}_A}{2}\right) \exp\left[\frac{ik}{L}(\mathbf{R}_I - \mathbf{R}_O)\cdot \mathbf{r}_A\right] g\left(\mathbf{R}_O, L; \mathbf{R}_A + \frac{\mathbf{r}_A}{2}, 0\right) g^*\left(\mathbf{R}_O, L; \mathbf{R}_A - \frac{\mathbf{r}_A}{2}, 0\right), \quad (8)$$

bears no dependence on the background and contrast of the object. It is expected to approach ±1 away from the edge, and all turbulence-induced variations are concentrated along the edge. It crucial to note that both the background and contrast values are measured directly from the individual image. This was be used in the measurements reported further in this paper.

In the following two sections we consider the variance of the normalized image fluctuations,

$$\sigma^2(X_I) = \langle i^2(X_I) \rangle - \langle i(X_I) \rangle^2. \quad (9)$$

Note also that traditional in beam scintillation studies normalization of the irradiance variance by mean irradiance, which leads to the scintillation index, is not practical for the edge image.

### 3. Phase screen model

#### 3.1 General formulation and mean image

As an introductory model we consider image distortions caused by a phase screen located at the imaging aperture. In this case factor $g(\mathbf{r}_0,0;\mathbf{r},z)$ in Eq. (1) is formulated as

$$g(\mathbf{R}_A, L; \mathbf{R}_A, 0) = \exp[iS(\mathbf{R}_A)]. \quad (10)$$

Here $S(\mathbf{R}_A)$ is the phase perturbation in the aperture plane, which replaces the extended turbulence medium between the object and aperture plane. Under standard assumptions of zero mean Gaussian phase with structure function

$$\left\langle \left[S\left(\mathbf{R}_A + \frac{\mathbf{r}_A}{2}\right) - S\left(\mathbf{R}_A - \frac{\mathbf{r}_A}{2}\right)\right]^2 \right\rangle = D(\mathbf{r}_A) \quad (11)$$

we calculate the mean image normalized image as

$$\langle i(X_I) \rangle = \frac{k}{2\pi L \Sigma_A} \int dX_O \, sign(X_O) \iint dX_A dY_A \int dx_A A\left(X_A + \frac{x_A}{2}, Y_A\right) A\left(X_A - \frac{x_A}{2}, Y_A\right)$$
$$\times \exp\left[\frac{ik}{L}(X_I - X_O)x_A - \frac{1}{2}D(x_A,0)\right]. \quad (12)$$

As expected, mean image is uniform in Y, and 1-D spatial spectrum of mean image

$$\langle \hat{i}(p) \rangle \equiv \frac{1}{2\pi} \int dX_I \langle i(X_I) \rangle \exp(-ipX_I) = \left[ \frac{1}{2\pi} \int dX_O \mathrm{sign}(X_O) \exp(-iX_O p) \right]$$
$$\times \left[ \frac{1}{\Sigma_A} \iint dX_A dY_A A\left(X_A + \frac{pL}{2k}, Y_A\right) A\left(X_A - \frac{pL}{2k}, Y_A\right) \right] \exp\left[-\frac{1}{2} D\left(\frac{pL}{k}, 0\right)\right] \quad (13)$$

is the product of the object spectrum, and aperture and turbulence filters, e. g. [11]. Integral over the object plane in Eq. (12) can be calculated with the help of the regularization technique, as shown in Appendix A, and Eq. (12) is simplified to

$$\langle i(X_I) \rangle = \frac{2}{\pi \Sigma_A} \int_0^\infty \frac{dx_A}{x_A} \sin\left(\frac{k}{L} X_I x_A\right) \exp\left[-\frac{1}{2} D(x_A, 0)\right] B_A(x_A)$$
$$B_A(x_A) \equiv \iint dX_A dY_A A\left(X_A + \frac{x_A}{2}, Y_A\right) A\left(X_A - \frac{x_A}{2}, Y_A\right). \quad (14)$$

As expected, mean image profile is odd with zero at the edge location. For circular aperture with radius $a$,

$$B_A(x_A) = 2a^2 \arccos\frac{x_A}{2a} - \frac{x_A}{2}\sqrt{4a^2 - x^2}, \quad (15)$$

and for power-law phase structure function

$$D(\mathbf{r}) = \left(\frac{r}{r_C}\right)^\alpha \quad (16)$$

equation for the mean image is presented as

$$\langle i(X_I) \rangle = \frac{4}{\pi^2} \int_0^1 \frac{dx}{x} \sin\left(\frac{2ak}{L} X_I x\right) \exp\left[-2^{\alpha-1}\left(\frac{a}{r_C}\right)^\alpha x^\alpha\right] \left(\arccos x - x\sqrt{1-x^2}\right) \quad (17)$$

In Fig. 2 mean image profiles with width normalize by diffraction limited width $L/ka$ are shown for $\alpha = 5/3$, and different values of parameter $a/r_C$. The profile width increases for large values of the $a/r_C$ ratio, and for $a/r_C > 1$ the mean profile can be presented as

$$\langle i(X_I) \rangle \approx \frac{2}{\pi} \int_0^\infty \frac{dy}{y} \sin\left(\frac{2kr_C}{L} X_I y\right) \exp(-2^{\alpha-1} y^\alpha), \quad (18)$$

suggesting that in this case the effective width is scaled by diffraction on the coherence radius scale $L/kr_C$.

### 3.2 Variance of normalized images. Perturbation Theory.

Using Eqs. (10), (11) in Eq. (8) one can calculate the second statistical moment of normalized image

$$\langle i^2(X_I) \rangle = \frac{k^2}{4\pi^2 L^2 \Sigma_A^2} \iint dX_{A1} dY_{A1} \int dx_{A1} A\left(X_{A1} + \frac{x_{A1}}{2}, Y_{A1}\right) A\left(X_{A1} - \frac{x_{A1}}{2}, Y_{A1}\right)$$
$$\times \int dX_{O1} \mathrm{sign}(X_{O1}) \exp\left(\frac{ik}{L}(X_I - X_{O1})x_{A1}\right) \iint dX_{A2} dY_{A2} \int dx_{A2} A\left(X_{A2} + \frac{x_{A2}}{2}, Y_{A2}\right) A\left(X_{A2} - \frac{x_{A2}}{2}, Y_{A2}\right) \quad (19)$$
$$\times \int dX_{O2} \mathrm{sign}(X_{O2}) \exp\left(\frac{ik}{L}(X_I - X_{O2})x_{A2}\right) \exp\left[-\frac{1}{2} \Psi(X_{A1}, Y_{A1}, x_{A1}, X_{A2}, Y_{A2}, x_{A2})\right].$$

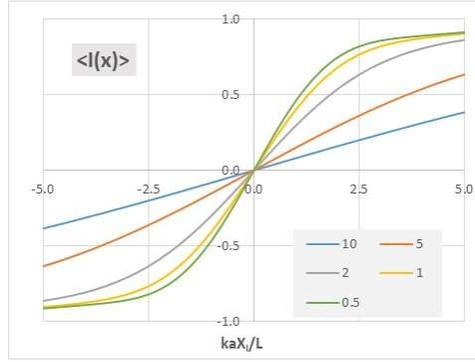

Fig. 2. Mean image profiles as functions of normalized coordinate. Parameter is $a/r_C$ ratio.

Here

$$\Psi(X_{A1},Y_{A1},x_{A1},X_{A2},Y_{A2},x_{A2}) = D(x_{A1},0) + D(x_{A2},0) + D\left(X_{A1}-X_{A2}+\frac{x_{A1}-x_{A2}}{2}, Y_{A1}-Y_{A2}\right)$$
$$+ D\left(X_{A1}-X_{A2}-\frac{x_{A1}-x_{A2}}{2}, Y_{A1}-Y_{A2}\right) - D\left(X_{A1}-X_{A2}+\frac{x_{A1}+x_{A2}}{2}, Y_{A1}-Y_{A2}\right) \qquad (20)$$
$$- D\left(X_{A1}-X_{A2}-\frac{x_{A1}+x_{A2}}{2}, Y_{A1}-Y_{A2}\right).$$

Eq. (19) is too complex for direct numerical evaluation, and we consider two asymptotic cases: weak and strong image blur. For an aperture with radius $a$, and power-law phase structure function with coherence radius $r_C$, the weak fluctuation case corresponds to $r_C > a$, and Eq. (19) can be evaluated using perturbation theory (PT)

$$\exp\left(-\frac{1}{2}\Psi\right) \approx 1 - \frac{1}{2}\Psi + \ldots \qquad (21)$$

For the image variance this leads to

$$\sigma_{PT}^2(X_I) = \left\langle i^2(X_I)\right\rangle_{PT} - \left\langle i(X_I)\right\rangle^2 = \iint dpdq\, \Phi_S(p,q)|P(p,q,X_I)|^2, \qquad (22)$$

where we introduced an auxiliary function

$$P(p,q,X_I) \equiv \frac{k}{\pi L \Sigma_A} \iint dX_A dY_A \int dx_A A\left(X_A+\frac{x_A}{2},Y_A\right) A\left(X_A-\frac{x_A}{2},Y_A\right)$$
$$\times \int dX_O \operatorname{sign}(X_O)\exp\left(\frac{ik}{L}(X_I-X_O)x_A + ipX_A + iqY_A\right)\sin\left(\frac{1}{2}px_A\right), \qquad (23)$$

and used spectral representation of structure function

$$D(x_A,y_A) = 2\iint dpdq\, \Phi_S(p,q)[1-\exp(ipx_A+iqy_A)]. \qquad (24)$$

Integral over the object plane can be calculated, as shown in Appendix A, resulting in

$$P(p,q,X_I) = \frac{4}{\pi \Sigma_A}\int_0^\infty \frac{dx_A}{x_A}\sin\left(\frac{1}{2}px_A\right)\cos\left(\frac{k}{L}X_I x_A\right)$$
$$\times \iint dX_A dY_A A\left(X_A+\frac{x_A}{2},Y_A\right) A\left(X_A-\frac{x_A}{2},Y_A\right)\exp(ipX_A+iqY_A). \qquad (25)$$

In case of the power-law structure function, Eq. (16), phase spectrum is

$$\Phi_S(p,q) = C_S(\alpha) r_C^{-\alpha} \left(p^2 + q^2\right)^{-1-\alpha/2}, \quad C_S(\alpha) = \frac{\alpha 2^{\alpha-2} \Gamma\left(1+\frac{\alpha}{2}\right)}{\pi \Gamma\left(1-\frac{\alpha}{2}\right)}, \tag{26}$$

and for a circular aperture with radius $a$ image variance is presented as

$$\sigma_{PT}^2(X_I) = C_{PT}\left(\frac{ka}{L} X_I\right)\left(\frac{a}{r_C}\right)^{\alpha} = C_{PT}\left(\frac{ka}{L} X_I\right) q^{-\alpha/2} N^{\alpha/2}. \tag{27}$$

Here

$$C_{PT}(\xi) = C_S(\alpha) \iint dp\,dq \left(p^2 + q^2\right)^{1-\alpha/2} \left|\tilde{P}(p,q,\xi)\right|^2,$$

$$\tilde{P}(p,q,\xi) = \frac{16}{\pi^2} \int_0^\infty \frac{dy}{y} \sin(py)\cos(2\xi y) \int_0^{1-y} dx \cos(px) \frac{\sin\left(q\sqrt{1-(x+y)^2}\right)}{q}, \tag{28}$$

and we introduced dimensionless parameters – Fresnel numbers for aperture size and coherence radius

$$N = \frac{ka^2}{L}, \quad q = \frac{kr_C^2}{L}. \tag{29}$$

Parameter $q$ is related to the traditional "Rytov variance" $\beta_0^2$. Specifically, for $\alpha = 5/3$

$$q \propto \left(\beta_C^2\right)^{-6/5}. \tag{30}$$

Formally, Eq. (28) presents image variance as a six-fold integral, but the special structure of this equation allows for a reasonably accurate numerical evaluation on a standard PC. We used a rings partition of the annulus at the $(p,q)$ plane with wave numbers range from $10^{-4}$ to $10^2$ in up to 24 log-uniform wide rings, and created from 20x20 to 60x60 uniform in angle and wavenumber grids in each ring. For each $(p,q)$ pair the non-singular double integral for $\tilde{P}(p,q,X_I)$ was calculated using MATLAB *integral2* function, and used in the corresponding integral sum. The low-wavenumber contribution was calculated analytically and added to the integral sum over the annulus.

The PT variance is concentrated along the image edge, and the width of the fluctuating area is of the order of diffraction-limited resolution $L/ka$. The variance is greatest at the edge location, $X_I = 0$, and for Kolmogorov case, $\alpha = 5/3$, the edge variance is calculated as

$$\sigma_{PT}^2(0) = 0.240 \left(\frac{a}{r_C}\right)^{5/3}. \tag{31}$$

Fig 3 shows PT variance profile, $C_{PT}(\xi)$, Eq. (27) as a function of normalized image plane coordinate $\xi = kaX_I/L$ for $\alpha = 5/3$. This profile is well approximated by a Gaussian curve

$$C_{PT}(\xi) = C_{PT}(0)\exp\left[-\left(\frac{\xi}{1.34}\right)^2\right]. \tag{32}$$

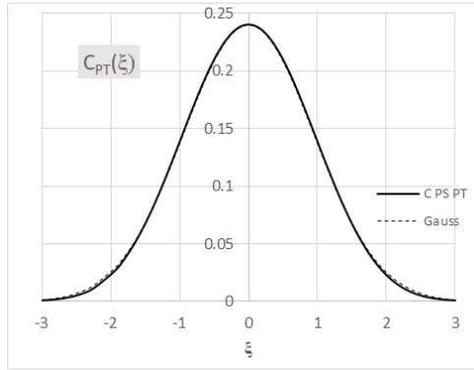

Fig. 3. Profile of PT image variance as function of $\xi = kaX_I/L$

### 3.3 Variance of normalized images. Coherence channels.

Strong fluctuation case can be investigated using Coherence Channels (CC) expansion. This technique was first described in [14], and more recently in [15, 16]. Essentially, the $\exp\left(-\frac{1}{2}\Psi\right)$ factor in Eq. (19) is approximated as sum of contributions of two domains where $\Psi < 1$

$$\exp\left[-\frac{1}{2}\Psi(X_{A1},Y_{A1},x_{A1},X_{A2},Y_{A2},x_{A2})\right] = M_0 + M_1 + ... + A_0 + A_1 + ... \qquad (33)$$

Here

$$M_0 = \exp\left[-\frac{1}{2}D(x_{A1},0) - \frac{1}{2}D(x_{A2},0)\right],$$

$$M_1 = -\frac{1}{2}M_0 \left[ \begin{array}{l} D\left(X_{A1}-X_{A2}+\frac{x_{A1}-x_{A2}}{2}, Y_{A1}-Y_{A2}\right) + D\left(X_{A1}-X_{A2}-\frac{x_{A1}-x_{A2}}{2}, Y_{A1}-Y_{A2}\right) \\ -D\left(X_{A1}-X_{A2}+\frac{x_{A1}+x_{A2}}{2}, Y_{A1}-Y_{A2}\right) - D\left(X_{A1}-X_{A2}-\frac{x_{A1}+x_{A2}}{2}, Y_{A1}-Y_{A2}\right) \end{array} \right] \qquad (34)$$

corresponds to the Main Coherence Channel, MCC, and

$$A_0 = \exp\left[-\frac{1}{2}D\left(X_{A1}-X_{A2}+\frac{x_{A1}-x_{A2}}{2}, Y_{A1}-Y_{A2}\right) - \frac{1}{2}D\left(X_{A1}-X_{A2}-\frac{x_{A1}-x_{A2}}{2}, Y_{A1}-Y_{A2}\right)\right]$$

$$A_1 = -\frac{1}{2}A_0 \left[ \begin{array}{l} D(x_{A1},0) + D(x_{A2},0) - -D\left(X_{A1}-X_{A2}+\frac{x_{A1}+x_{A2}}{2}, Y_{A1}-Y_{A2}\right) \\ -D\left(X_{A1}-X_{A2}-\frac{x_{A1}+x_{A2}}{2}, Y_{A1}-Y_{A2}\right) \end{array} \right] \qquad (35)$$

is the Additional Coherence Channel (ACC).

The $M_0$ term is just $\langle i(X_I) \rangle^2$, and makes no contribution to image variance. The first-order MCC term is calculated using the necessary CC condition $a > r_C$, and regularization of the object plane integral described in Appendix as

$$\sigma_{M1}^2(X_I) = \iint dp\,dq\,\Phi(p,q)|M(p,q,X_I)|^2, \qquad (36)$$

where we introduced an auxiliary function

$$M(p,q,X_I) = \frac{2}{\pi\Sigma_A} p \int_0^\infty dx_A \cos\left(\frac{k}{L}X_I x_A\right) \exp\left[-\frac{1}{2}D(x_A)\right] \qquad (37)$$
$$\times \iint dX_A dY_A A^2(X_A, Y_A) \exp(ipX_A + iqY_A).$$

For a circular aperture with radius $a$, and power-law structure function, Eq. (16), the MCC contribution to variance is calculated from Eqs. (36, 37) as

$$\sigma^2(X_I)_{M1} = C_{M1}\left(\frac{kr_C}{L}X_I\right)\left(\frac{a}{r_C}\right)^{\alpha-2}, \quad C_{M1}(\eta) = C_S(\alpha)\frac{16}{\pi^2}T_1 T_2^2(\eta), \qquad (38)$$

where

$$T_1 = \iint dp\, dq\, p^2 (p^2+q^2)^{-2-\alpha/2} J_1^2(p^2+q^2) = \sqrt{\pi}\,\frac{\Gamma\left(1-\frac{\alpha}{2}\right)\Gamma\left(\frac{1+\alpha}{2}\right)}{2\left(1+\frac{\alpha}{2}\right)\Gamma^2\left(1+\frac{\alpha}{2}\right)}, \qquad (39)$$

$$T_2(\eta) = \int_0^\infty dt\, \cos(\eta t)\exp\left(-\frac{1}{2}t^\alpha\right).$$

Contribution of the $A_0$ term of the CC expansion, Eq. (29) can be estimated as

$$\sigma_{A0}^2(X_I) = O\left(\frac{r_C^2}{a^2}\right). \qquad (40)$$

Under condition $a > r_C$ required for the CC expansion, the $A_0$ term is smaller than the $M_1$ term, Eq. (38). The $A_1$ term is expected to be even smaller in the CC validity domain. Since the $A_0$ term is smaller than the $M_1$ term when $a > r_C$, we conclude that the $M_1$ term, Eq. (38), is the leading term of the CC series, and will be assigned a CC subscript further on.

For Kolmogorov case, $\alpha = 5/3$ we have $T_1 = 2.7155$, $T_2(0) = 1.3543$, and edge CC variance is

$$\sigma_{CC}^2(0) = 0.575\left(\frac{a}{r_C}\right)^{-1/3}. \qquad (41)$$

The width of the variance profile, as suggested by Eq. (38), is scaled by diffraction on the coherence radius $L/kr_C$.

Fig. 4 shows CC variance profile, $C_{CC}(\eta)$, Eq. (38) as a function of normalized image plane coordinate $\eta = (kr_C/L)X_I$ for $\alpha = 5/3$.

Two asymptotes for the image variance, Eq. (27) and Eq. (36) are based on complementary assumptions: $a < r_C$, and $a > r_C$. It is essential, that both PT, Eq. (31) and CC, Eq. (41), results for the edge variance are of the order of unity for $a \propto r_C$. Additionally, the effective widths of the PT and CC variance distributions, $L/ka$ and $L/kr_C$ also are f the same order when $a \propto r_C$. Finally, we conclude that the PT and CC results for a complete set of asymptote for the image variance for the phase screen imaging model.

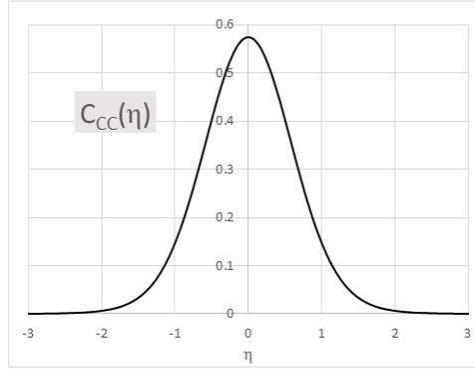

Fig. 4. Profile of CC image variance as function of $\eta = (kr_C/L)X_I$

Fig. (5) shows the PT and CC results, Eq. (31) and Eq. (41) for the edge variances. Two asymptotes intersect at $a \approx 1.5 r_C$, at the $\sigma^2(0) \approx 0.5$. While, keeping with traditional designation, we used "strong fluctuation" term, in fact image variance is always smaller than unity, and even decreases for lager $a/r_C$ values. By the very nature of asymptotic theories, these results are expected to be accurate for very small and very large values of the $a/r_C$ ratio, and no rigorous variance estimates can be made for $a \approx r_C$. However, invoking physically reasonable assumption that the variance dependence is a smooth function of $a/r_C$, one can offer some simple practical uniform approximation for dependence of edge variance on $a/r_C$. Namely, following some examples in [17], we introduce a set of uniform approximations in the following form related to harmonic mean of PT and CC results

$$\sigma_{HM}^2(0) = \left[ \left(\sigma_{PT}^2(0)\right)^{-n} + \left(\sigma_{CC}^2(0)\right)^{-n} \right]^{-1/n}. \tag{42}$$

Fig. 5 shows examples of uniform approximations, Eq. (42) for $n = 1, 2$. By design, smooth approximations are close to the rigorous asymptotes for large and small $a/r_C$ values, and reach maxima at $a \approx r_C$. The actual edge variance can be expected to be in the 0.3 – 0.5 range.

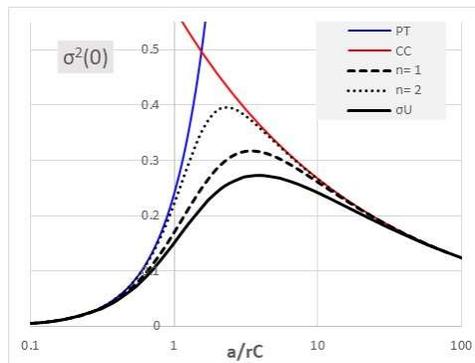

Fig. 5. PT and CC asymptotes, Eq. (31) and Eq. (41), and HM approximations, Eq. (42) (dashed curves) for $n = 1, 2$. Heavy solid curve – uniform approximation, Eq. (39, 40).

### 3.4 Uniform approximation

While smooth merging of the edge variance, $\sigma^2(0)$, asymptotes is fairly straightforward, designing a uniform approximation for variance profiles, $\sigma^2(X_I)$ is a more challenging task.

Fortunately, the analytical forms of the PT and CC asymptotes, Eq. (22) and Eq. (36) suggest that a uniform approximation can be sought as

$$\sigma_U^2(X_I) = \iint dp\,dq\,\Phi(p,q)|U(p,q,X_I)|^2, \quad (43)$$

where

$$U(p,q,X_I) = \frac{4}{\pi\Sigma_A}\int_0^\infty \frac{dx_A}{x_A}\sin\left(\frac{1}{2}px_A\right)\cos\left(\frac{k}{L}X_I x_A\right)\exp\left[-\frac{1}{2}D(x_A)\right]$$
$$\times \iint dX_A\,dY_A\,A\left(X_A + \frac{x_A}{2}, Y_A\right)A\left(X_A - \frac{x_A}{2}, Y_A\right)\exp(ipX_A + iqY_A). \quad (44)$$

It is straightforward to check that Eq. (44) reduces to Eq. (23) for $a < r_C$, and to Eq. (37) for $a > r_C$. Eq. (43) and Eq. (44) describe smooth transition both of the edge variance and image variance profiles from PT to CC asymptotes. Further transformations for circular apertures and power-law structure function are similar to ones leading to Eqs. (27, 28). Numerical calculations of $\sigma_U^2(X_I)$ are possible by integration procedure described for the PT case. Dot-dashed curve in Fig. 4 shows edge variance $\sigma_U^2(0)$. This curve is similar to the more formal harmonic approximations, but has slightly smaller maximum.

Fig. 6 shows normalized profiles $\sigma_U^2(X_I)/\sigma_U^2(0)$ as functions of normalized image plane coordinate, $\xi = (2ka/L)X_I$. Profile width gradually increases from diffraction limit, Fig. 2, in the PT case, $a < r_C$, to one scaled by diffraction on coherence radius for the CC case, $a > r_C$. Fig. 7 shows normalized profiles $\sigma_U^2(X_I)/\sigma_U^2(0)$ as functions of normalized image space coordinate $\eta = (kr_C/L)X_I$. For $a \geq 3r_C$ profiles are close to the CC limit shown in Fig. 4.

Finally, we emphasize that Eqs. (43, 44) do not represent an exact image variance solution, but only an approximation that is accurate only for $a > r_C$ and $a < r_C$.

## 4. Extended medium

### 4.1 Path integral formulation

In the case of extended turbulent medium between the source and the aperture in the form of Feynman path integral [10, 15]. We refer to [9] for the details of derivation, and display only the essential steps here. Perturbation factor $g(\mathbf{r}_0,0;\mathbf{r},z)$ in Eq. (1) can be presented as

$$g(\mathbf{r}_0,0;\mathbf{r},z) = \frac{2\pi i|z|}{k}\iint D\mathbf{v}(\zeta)\delta\left(\int_0^z d\zeta\,\mathbf{v}(\zeta)\right)$$
$$\exp\left\{\frac{ik}{2}\int_0^z d\zeta\,v^2(\zeta) + ik\int_0^z d\zeta\,n\left(\mathbf{r}_0\left(1-\frac{\zeta}{z}\right) + \mathbf{r}\frac{\zeta}{z} + \tilde{\mathbf{r}}(\zeta),\zeta\right)\right\}. \quad (45)$$

Here $D\mathbf{v}(z)$ denotes integration over the functional space of two-dimensional continuous paths, $n(\mathbf{r},\zeta)$ is the random zero-average component of inhomogeneous refractive index, and

$$\tilde{\mathbf{r}}(\zeta) \equiv \int_\zeta^z d\zeta'\,\mathbf{v}(\zeta') \quad (46)$$

is the deviation of the Feynman path from the straight line.

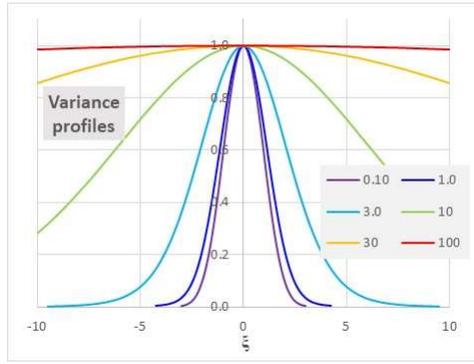

Fig. 6. Normalized profiles, $\sigma_U^2(X_I)/\sigma_U^2(0)$, as function of normalized coordinate $\xi = (ka/L)X_I$.

Normalized instantaneous image of the edge, Eq. (8) is now presented as

$$i(X_I,Y_I) = \frac{1}{\Sigma_A}\iint d^2R_O sign(X_O)\iint d^2R_A \iint d^2r_A A\left(\mathbf{R}_A + \frac{\mathbf{r}_A}{2}\right)A\left(\mathbf{R}_A - \frac{\mathbf{r}_A}{2}\right)$$

$$\times \exp\left[\frac{ik}{L}(\mathbf{R}_I - \mathbf{R}_O)\cdot\mathbf{r}_A\right]\iiint D\mathbf{V}(\zeta)\iint D\mathbf{v}(\zeta)\delta\left(\int_0^L d\zeta \mathbf{V}(\zeta)\right)\delta\left(\int_0^L d\zeta \mathbf{v}(\zeta)\right) \quad (47)$$

$$\times \exp\left\{\begin{array}{l}ik\int_0^L d\zeta\, \mathbf{V}(\zeta)\cdot\mathbf{v}(\zeta) + ik\int_0^L dz\, n\left(\mathbf{R}_0\frac{z}{L} + \left(\mathbf{R}_A + \frac{\mathbf{r}_A}{2}\right)\left(1-\frac{z}{L}\right) + \tilde{\mathbf{r}}_+(z),z\right) \\ -ik\int_0^L dz\, n\left(\mathbf{R}_0\frac{z}{L} + \left(\mathbf{R}_A - \frac{\mathbf{r}_A}{2}\right)\left(1-\frac{z}{L}\right) + \tilde{\mathbf{r}}_-(z),z\right)\end{array}\right\},$$

where

$$\tilde{\mathbf{r}}_\pm(z) = \int_z^L d\zeta\left[\mathbf{V}(\zeta) \pm \frac{1}{2}\mathbf{v}(\zeta)\right]. \quad (48)$$

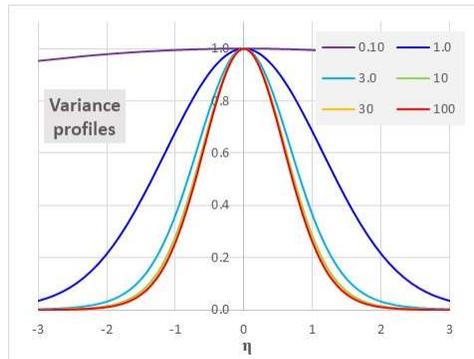

Fig. 7. Normalized profiles, $\sigma_U^2(X_I)/\sigma_U^2(0)$, as function of normalized coordinate $\eta = (kr_C/L)X_I$.

Calculations of statistical moments is performed for zero-mean normal random field of refractive index using Markov approximation [18]

$$\langle n(\mathbf{r}_1, z_1) n(\mathbf{r}_2, z_2) \rangle = A(\mathbf{r}_1 - \mathbf{r}_2, z_1) \delta(z_1 - z_2). \tag{49}$$

Here for statistically homogeneous propagation path

$$A(\mathbf{r}) = 2\pi \iint d^2\kappa \, \Phi_n(\mathbf{\kappa}, 0) \exp(i\mathbf{\kappa} \cdot \mathbf{r}). \tag{50}$$

In particular, mean image is calculated as [9]

$$\langle i(X_I) \rangle = \frac{k}{2\pi L \Sigma_A} \iiint dX_O \, \text{sign}(X_O) \iint dX_A dY_A \int dx_A A\left(X_A + \frac{x_A}{2}, Y_A\right) A\left(X_A - \frac{x_A}{2}, Y_A\right)$$
$$\times \exp\left[\frac{ik}{L}(X_I - X_O)x_A - \frac{\pi k^2}{4} \int_0^L dz \, H\left(x_A\left(1 - \frac{z}{L}\right)\right)\right]. \tag{51}$$

Here we used traditional [18] shorthand notation

$$H(\mathbf{r}) = 8 \iint d^2\kappa \, \Phi_n(\mathbf{\kappa}, 0)[1 - \exp(i\mathbf{\kappa} \cdot \mathbf{r})]. \tag{52}$$

Comparison of the extended medium average image, Eq. (51) to the phase screen result, Eq. (12) reveals that they are identical when

$$D(x_A) = \frac{\pi k^2}{2} \int_0^L dz \, H\left(x_A\left(1 - \frac{z}{L}\right)\right). \tag{53}$$

This is expected, as discussed in [13, 19]. In case of power-law spectrum, when

$$\Phi_n(K) = C_\Phi(\alpha) C_n^2 K^{-2-\alpha}, \quad C_\Phi(\alpha) = \frac{\Gamma(\alpha+1)}{4\pi^2} \sin\left(\frac{\pi}{2}(\alpha-1)\right),$$
$$H(r) = C_H(\alpha) C_n^2 r^\alpha, \quad C_H(\alpha) = C_\Phi(\alpha) \frac{16\pi \Gamma\left(1 - \frac{\alpha}{2}\right)}{\alpha 2^\alpha \Gamma\left(1 + \frac{\alpha}{2}\right)}, \tag{54}$$

We can relate coherence radius $r_C$ of the phase screen model to the turbulent path parameters as follows

$$r_C = \left(\frac{\pi C_H(\alpha)}{2(\alpha+1)} C_n^2 k^2 L\right)^{-1/\alpha} = \left(\frac{\pi^2 \Gamma\left(1 - \frac{\alpha}{2}\right) C_\Phi(\alpha)}{\alpha(\alpha+1) 2^{\alpha-3} \Gamma\left(1 + \frac{\alpha}{2}\right)} C_n^2 k^2 L\right)^{-1/\alpha} = \left(\frac{2\pi C_\Phi(\alpha)}{(\alpha+1) C_S(\alpha)} C_n^2 k^2 L\right)^{-1/\alpha}$$
. (55)

In Kolmogorov case, $\alpha = 5/3$, we calculate $C_\Phi(\alpha) = 0.033$, and $C_H(\alpha) = 1.855$. Coherence radius is related to the turbulence parameters as

$$r_C = 0.948 \left(C_n^2 k^2 L\right)^{-3/5} \tag{56}$$

To summarize, mean irradiance for the extended medium is still described by Eq. (17) provided that $r_C$ is given in Eq. (55).

Second moment of image is calculated based on Eq. (47) using Markov approximation, Eq. (49). We refer for details to [9]

$$\langle i^2(\mathbf{R}_I)\rangle = \frac{k^2}{4\pi^2 L^2 \Sigma_A^2} \iint D^2\mathbf{v}_1(\cdot) \iint D^2\mathbf{v}_2(\cdot) \exp\left(ik\int_0^L dz\,\mathbf{v}_1(z)\cdot\mathbf{v}_2(z)\right)\delta\left(\int_0^L dz\mathbf{v}_1(z)\right)\delta\left(\int_0^L dz\mathbf{v}_2(z)\right)$$

$$\times \iint d^2 R_{O1}\,\text{sign}(X_{O1}) \iint d^2 R_{A1}\int d^2 r_{A1} A\left(\mathbf{R}_{A1}+\frac{\mathbf{r}_{A1}}{2}\right) A\left(\mathbf{R}_{A1}-\frac{\mathbf{r}_{A1}}{2}\right)\exp\left(\frac{ik}{L}(\mathbf{R}_I-\mathbf{R}_{O1})\mathbf{r}_{A1}\right)$$

$$\times \iint d^2 R_{O2}\,\text{sign}(X_{O2}) \iint d^2 R_{A2}\int d^2 r_{A2} A\left(\mathbf{R}_{A2}+\frac{\mathbf{r}_{A2}}{2}\right) A\left(\mathbf{R}_{A2}-\frac{\mathbf{r}_{A2}}{2}\right)\exp\left(\frac{ik}{L}(\mathbf{R}_I-\mathbf{R}_{O2})\mathbf{r}_{A2}\right) \quad (57)$$

$$\times \exp\left[-\frac{\pi k^2}{4}\int_0^L dz\, F\bigl(\tilde{\mathbf{r}}_1(z),\tilde{\mathbf{r}}_1'(z),\tilde{\mathbf{r}}_2(z),\tilde{\mathbf{r}}_2'(z)\bigr)\right].$$

Here

$$F(\tilde{\mathbf{r}}_1,\tilde{\mathbf{r}}_1',\tilde{\mathbf{r}}_2,\tilde{\mathbf{r}}_2') = H(\tilde{\mathbf{r}}_1-\tilde{\mathbf{r}}_1') + H(\tilde{\mathbf{r}}_2-\tilde{\mathbf{r}}_2') + H(\tilde{\mathbf{r}}_1-\tilde{\mathbf{r}}_2') + H(\tilde{\mathbf{r}}_1'-\tilde{\mathbf{r}}_2) - H(\tilde{\mathbf{r}}_1-\tilde{\mathbf{r}}_2) - H(\tilde{\mathbf{r}}_1'-\tilde{\mathbf{r}}_2'), \quad (58)$$

and

$$\tilde{\mathbf{r}}_1(z) = \mathbf{R}_{O1}\frac{z}{L} + \left(\mathbf{R}_{A1}+\frac{\mathbf{r}_{A1}}{2}\right)\left(1-\frac{z}{L}\right) + \frac{1}{2}\int_z^L d\zeta\,[\mathbf{v}_1(\zeta)+\mathbf{v}_2(\zeta)],$$

$$\tilde{\mathbf{r}}_1'(z) = \mathbf{R}_{O1}\frac{z}{L} + \left(\mathbf{R}_{A1}-\frac{\mathbf{r}_{A1}}{2}\right)\left(1-\frac{z}{L}\right) + \frac{1}{2}\int_z^L d\zeta\,[\mathbf{v}_1(\zeta)-\mathbf{v}_2(\zeta)], \quad (59)$$

$$\tilde{\mathbf{r}}_2(z) = \mathbf{R}_{O2}\frac{z}{L} + \left(\mathbf{R}_{A2}+\frac{\mathbf{r}_{A2}}{2}\right)\left(1-\frac{z}{L}\right) - \frac{1}{2}\int_z^L d\zeta\,[\mathbf{v}_1(\zeta)+\mathbf{v}_2(\zeta)],$$

$$\tilde{\mathbf{r}}_2'(z) = \mathbf{R}_{O2}\frac{z}{L} + \left(\mathbf{R}_{A2}-\frac{\mathbf{r}_{A2}}{2}\right)\left(1-\frac{z}{L}\right) - \frac{1}{2}\int_z^L d\zeta\,[\mathbf{v}_1(\zeta)-\mathbf{v}_2(\zeta)]$$

Eq. (57) is clearly more complicated than the phase screen analog, Eq. (19). Besides the obvious path-integral presence, there are non-isoplanatic components in Eq. (59) that complicate calculations.

Similar to the phase screen case our further development is based on the asymptotic analysis where we consider PT and CC expansions. Unlike the phase screen case, due to the presence of path integrals, we are not able to explicitly state in advance the validity conditions for either approximation. We perform the expansions formally, and determine the validity regions based on comparison of the main terms of expansions.

### 4.2 Weak fluctuations. PT

For the PT case we use the first two orders of Taylor expansion of the last exponent in Eq. (57), which is similar to Eq. (21). As shown in Appendix B, the PT result for the image variance is presented as

$$\sigma_{PT}^2(X_I) = 2\pi k^2 \int_0^L dz \iint dp\,dq\,\Phi_n(p,q)|P(p,q,z,X_I)|^2, \quad (60)$$

where

$$P(p,q,z,X_I) = \frac{2}{\pi \Sigma_A} \int_{-\infty}^{\infty} \frac{dx}{\left(x - \frac{pz}{k}\right)} \sin\left[\frac{1}{2} p\left(x - \frac{pz}{k}\right)\left(1 - \frac{z}{L}\right)\right] \exp\left[\frac{ik}{L} X_I x\right]$$

$$\times \iint dX dY A\left(X + \frac{x}{2}, Y + \frac{qz}{2k}\right) A\left(X - \frac{x}{2}, Y - \frac{qz}{2k}\right) \exp\left[i(pX + qY)\left(1 - \frac{z}{L}\right)\right]. \quad (61)$$

It is important to note that variance profile is even in $X_I$, and the profile width is scaled as diffraction on the aperture size, $L/ka$. Eq. (61) has two asymptotic cases depending on the value of aperture Fresnel number, $N \equiv ka^2/L$. For $N > 1$ effective integration range for variable $x$ is of order of the aperture size, and Eq. (61) simplifies to

$$P_{N>1}(p,q,z,X_I)_{N>1} = \frac{4}{\pi \Sigma_A} \int_0^{\infty} \frac{dx}{x} \sin\left[\frac{1}{2} px\left(1 - \frac{z}{L}\right)\right] \cos\left(\frac{k}{L} X_I x\right)$$

$$\times \iint dX dY A\left(X + \frac{x}{2}, Y\right) A\left(X - \frac{x}{2}, Y\right) \exp\left[i(pX + qY)\left(1 - \frac{z}{L}\right)\right]. \quad (62)$$

In case of power-law refractive index spectra, Eq. (54), the $N > 1$ case is presented as

$$\sigma^2_{PT}(X_I)_{N>1} = \frac{2\pi C_\Phi(\alpha)}{\alpha + 1} C_n^2 k^2 L \iint dp dq (p^2 + q^2)^{-1-\alpha/2}$$

$$\times \left| \frac{4}{\pi \Sigma_A} \int_0^{\infty} \frac{dx}{x} \sin\left[\frac{1}{2} px\right] \cos\left(\frac{k}{L} X_I x\right) \iint dX dY A\left(X + \frac{x}{2}, Y\right) A\left(X - \frac{x}{2}, Y\right) \exp[i(pX + qY)] \right|^2 \quad (63)$$

Inspection of Eq. (63) reveals that when coherence radius is related to the turbulence parameters as was calculated in Eq. (55) based on mean irradiance, the extended medium result for $N > 1$ is identical to the PT phase screen result, Eqs. (22, 25). This is true even for the statistically inhomogeneous case, when $C_n^2$ caries along the propagation path. For a top-hat aperture Eqs. (27, 28) and Fig. 3 are accurate for the extended turbulence PT image variance when $N > 1$.

It is convenient to use spectral form of aperture function

$$A(x,y) = \iint ds dt \hat{A}(s,t) \exp(isx + iqy) \quad (64)$$

to represent Eq. (61) with lesser integration order

$$P(p,q,z,X_I) = \frac{8\pi^2}{\Sigma_A} \int_{-\infty}^{\infty} dt \int_{-\frac{p}{2}\left(1-\frac{z}{L}\right)}^{\frac{p}{2}\left(1-\frac{z}{L}\right)} dt\, \hat{A}^*\left(\frac{kX_I}{L} + s + \frac{p}{2}\left(1 - \frac{z}{L}\right), t + \frac{q}{2}\left(1 - \frac{z}{L}\right)\right)$$

$$\times \hat{A}\left(\frac{kX_I}{L} + s - \frac{p}{2}\left(1 - \frac{z}{L}\right), t - \frac{q}{2}\left(1 - \frac{z}{L}\right)\right) \exp\left[\frac{iz}{k}(ps + qt)\right] \quad (65)$$

For $N < 1$, Eq. (65) can be further simplified to

$$P(p,q,z,X_I)_{N<1} = \frac{8\pi^2}{\Sigma_A} \left| p\left(1 - \frac{z}{L}\right) \right| \int_{-\infty}^{\infty} dt \left| \hat{A}^*\left(\frac{kX_I}{L}, t\right) \right|^2 \exp\left(\frac{iz}{k} qt\right), \quad (66)$$

and in case of the power-law spectra and top-hat aperture image variance can be further reduced to the two-fold integral

$$\sigma^2_{PT}(X_I)_{N<1} = 2\pi C_\Phi(\alpha) C_n^2 k^2 \frac{16}{\pi^2} \int_0^L dz \left(1 - \frac{z}{L}\right)^2 \iint dp\,dq\, p^2 (p^2 + q^2)^{-1-\alpha/2}$$

$$\times \left| \int_0^\infty \frac{dt}{\left(\frac{k^2 X_I^2}{L^2} + t^2\right)} J_1^2\left(a\sqrt{\frac{k^2 X_I^2}{L^2} + t^2}\right) \cos\left(\frac{z}{k}qt\right) \right|^2 \tag{67}$$

Eq. (67) can be further simplified as shown in Appendix C, and we use Eq. (C6) presented it as

$$\sigma^2_{PT}(X_I)_{N<1} = C_{PT}\left(\frac{ka}{L} X_I\right)_{N<1} r_C^{-\alpha} a^{4-\alpha} k^{2-\alpha} L^{\alpha-2} = C_{PT}\left(\frac{ka}{L} X_I\right)_{N<1} q^{-\alpha/2} N^{2-\alpha/2}. \tag{68}$$

where

$$C_{PT}(\xi)_{N<1} = \frac{2^{\alpha+2}}{\pi^2 \sqrt{\pi}} \frac{\Gamma\left(\frac{\alpha-1}{2}\right)}{\Gamma\left(1 - \frac{\alpha}{2}\right)} \frac{\Gamma(2-\alpha)}{(\alpha-1)} \cos\left(\frac{\pi(\alpha-1)}{2}\right) \int_0^\infty \frac{dy}{(\xi^2 + y^2)} J_1^2\left(\sqrt{\xi^2 + y^2}\right)$$

$$\times \int_0^\infty \frac{dy'}{(\xi^2 + y'^2)} J_1^2\left(\sqrt{\xi^2 + y'^2}\right) \left(|y + y'|^{\alpha-2} + |y - y'|^{\alpha-2}\right) \tag{69}$$

For $\alpha = 5/3$ the edge variance can be calculated numerically as

$$\sigma^2_{PT}(0)_{N<1} = 0.474\, r_C^{-5/3} a^{7/3} k^{1/3} L^{-1/3} = 0.474\, q^{-5/6} N^{7/6}. \tag{70}$$

Note that, unlike the $N > 1$ case, Eqs. (68, 69) are exact for the statistically uniform case only, and is only an order-of-magnitude correct for statistically inhomogeneous propagation paths. This result counterparts the $N > 1$ asymptote, Eq. (27), and evidently both results are of the same order at $N \propto 1$. Fig. 7 shows both PT asymptotes as function of aperture Fresnel number $N$ for $q = 1$. Asymptotes intersect at $N \approx 0.13$ where the edge variance growth rate reduces from $\sigma^2_{PT}(0) \propto a^{7/3}$ to $\sigma^2_{PT}(0) \propto a^{5/3}$. Integration of Eq. (69) for $\xi \neq 0$ uncovers that the shape of the normalized variance profile $C_{PT}(\xi)/C_{PT}(0)$ for $N < 1$ is indistinguishable from the $N > 1$ case in Fig. 3, and is well approximated by Gaussian curve, Eq. (32).

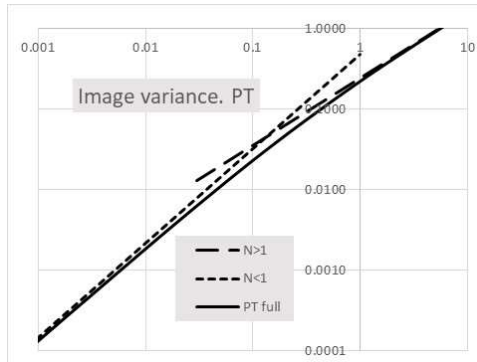

Fig. 7. Dependence of the image variance at the edge location on aperture Fresnel number $N$ calculated using PT for $q = 1$. Short-dash line - $N < 1$ case, Eq. (68), long-dash line - $N > 1$ case, Eq. (27). Solid curve – complete PT, Eqs. (60, 61).

Complete PT result, Eqs. (60, 61), is rather complicated, but is still amendable for numerical integration using procedure similar to one used for Eq. (28). Solid curve in Fig. 7 shows result of these calculations for edge variance that smoothly merge the $N<1$ and $N>1$ asymptotes. Given above mentioned similarity of the variance profiles for the $N<1$ and $N>1$ asymptotes, it is reasonable to assume that the same shape is valid also for $N \propto 1$ as well.

It is important to emphasize the fundamental difference between the results presented by solid curves at the phase screen Fig. (5) and PT case, Eq. (7). In the former case solid curve in Fig. 5, Eqs. (43, 44), is an ad hoc approximation, which accuracy cannot be improved by controlling the physical parameters of the problem. In contrast, the solid curve in Fig. 7 is a rigorous asymptotic result that becomes more accurate when turbulence is weaker.

As Fig. 7 indicate, for any turbulence strength, determined by parameter $q$, image variance growth unlimited when aperture size increases. Intuitively, this contradict the still undetermined PT restrictions. At the same time, it indicates that for the problem under consideration classic, $q>1$, condition of weak fluctuations is not valid.

### 4.3 Strong fluctuations. CC expansion

As an alternative to the PT expansion, we apply CC expansion, similar to Eq. (33), to the last exponent in Eq. (57). Similar to the phase screen case, the $M_0$ term of CC series is just a square of mean irradiance, and does not make a contribution to the image variance. We refer to [9] for details.

The path integrals in the $M_1$ term can be evaluated as described in Appendix D, and is presented as

$$\sigma_{M1}^2(X_I) = 2\pi k^2 \int_0^L dz \iint dpdq \Phi_n(p,q) |M(p,q,z,X_I)|^2, \tag{71}$$

where

$$M(p,q,z,X_I) = \frac{2}{\pi \Sigma_A} \int_{-\infty}^{\infty} \frac{dx}{x} \sin\left[\frac{1}{2}px\left(1-\frac{z}{L}\right)\right] \exp\left[\frac{ik}{L}X_I\left(x+\frac{pz}{k}\right) - \frac{\pi k^2}{4}\int_0^L d\zeta H[\mathbf{r}(z,\zeta)]\right]$$

$$\times \iint dXdY A\left(X+\frac{x}{2}+\frac{pz}{2k}, Y+\frac{qz}{2k}\right) A\left(X-\frac{x}{2}-\frac{pz}{2k}, Y-\frac{qz}{2k}\right) \exp\left[i(pX+qY)\left(1-\frac{z}{L}\right)\right]. \tag{72}$$

CC expansion requires that the turbulence-related exponent is effectively limit the integration domain. Inspection of Eq. (72) shows that this requires that $r_C < a$ or $q < N$, and Eq. (72) simplifies to

$$M(p,q,z,X_I) = \frac{2}{\pi \Sigma_A} \int_{-\infty}^{\infty} \frac{dx}{x} \sin\left[\frac{1}{2}px\left(1-\frac{z}{L}\right)\right] \exp\left(\frac{ik}{L}X_I\left(x+\frac{pz}{k}\right) - \frac{\pi k^2}{4}\int_0^L d\zeta H[\mathbf{r}(z,\zeta)]\right)$$

$$\times \iint dXdY A^2(X,Y) \exp\left[i(pX+qY)\left(1-\frac{z}{L}\right)\right]. \tag{73}$$

There are two asymptotic cases for Eq. (73). When $qN>1$, we have $x \propto r_C$, and $p,q \propto 1/a$, and Eq. (73) reduces to

$$M(p,q,z,X_I) = \frac{2}{\pi \Sigma_A} p\left(1-\frac{z}{L}\right) \int_0^\infty dx \cos\left(\frac{k}{L} X_I x\right) \exp\left(-\frac{\pi k^2}{4} \int_0^L d\zeta H\left[x\left(1-\frac{\zeta}{L}\right)\right]\right) \quad (74)$$
$$\times \iint dX dY A^2(X,Y) \exp\left[i(pX+qY)\left(1-\frac{z}{L}\right)\right].$$

Eq. (74) is very similar to the phase screen model Eq. (37). For power law spectra, Eq. (54), when coherence radius is related to the propagation path parameters by Eq. (55), the MCC contribution to variance is the same as given by Eqs. (37) for the phase screen model. For a circular aperture with radius $a$ it is specified by Eq. (38).

When $qN < 1$, we have $x \propto r_C$, and $p,q \propto kr_C/L$, and Eq. (73) reduces to

$$M(p,q,z,X_I) = \frac{1}{\pi} p\left(1-\frac{z}{L}\right) \int_{-\infty}^\infty dx \cos\left[\frac{k}{L} X_I\left(x+\frac{pz}{k}\right)\right] \exp\left(-\frac{\pi k^2}{4} \int_0^L d\zeta H[\mathbf{r}(z,\zeta)]\right). \quad (75)$$

Examination of Eq. (75) and Eq. (71) reveals that the image edge variance under conditions $q < N$ and $qN < 1$ bears no dependence on the aperture size, and is exclusively determined by turbulent propagation path.

$$\sigma_{M1}^2(0) = \frac{2\pi k^2}{\pi^2} \int_0^L dz \left(1-\frac{z}{L}\right)^2 \iint dp\, dq\, p^2 \Phi_n(p,q)$$
$$\times \left| \int_{-\infty}^\infty dx \exp\left(-\frac{\pi k^2}{4} \int_0^L d\zeta H\left[x\left(1-\frac{\zeta}{L}\right)+\frac{p}{k}[z-\min(z,\zeta)], \frac{q}{k}[z-\min(z,\zeta)]\right]\right) \right|^2. \quad (76)$$

It is straightforward to show that for power-law spectra, Eq. (54), in Eq. (76) $x \propto r_C$, and $p,q \propto kr_C/L$, and

$$\sigma_{M1}^2(0) = O\left((C_n^2)^{2-\frac{4}{\alpha}} k^{6-\alpha-\frac{8}{\alpha}} L^{\alpha-\frac{4}{\alpha}}\right) = O(q^{2-\alpha}). \quad (77)$$

Contribution of the zero-order ACC term is evaluated in Appendix E, where two asymptotic cases are given by Eq. (E6) and Eq. (E8). The corresponding MCC results are given by Eq. (38) and Eq. (77). Under conditions $q < N$ in both cases ACC tern is smaller than the MCC term, and can be neglected.

### 4.4 Review of asymptotes. Discussion

For power-law turbulence spectra, Eq. (54) edge scintillation index is a function of two dimensionless parameters: Fresnel numbers for the aperture size, $N$, and coherence radius, $q$. Our asymptotic analysis for the extended medium with power law spectra led to four distinct equation Eq. (31), Eq. (70), Eq. (41) and Eq. (77) for the edge image variance. It is important that the CC results, as derived, have a well-defined restriction, $q < N$. Additionally, CC has two subdomains: $qN > 1$ for Eq. (41) and $qN < 1$ for Eq. (77). Validity domains for the PT results, Eq. (31) and Eq. (70) have not been explicitly determined, aside from the intuitive notion that PT is valid for weak turbulence, namely for large $q$ values.

We found it instructive to present these asymptotes on the $(N,q)$ plane, similar to asymptote maps used in [15]. Fig. 8 shows four asymptotes for the image variance for $\alpha = 5/3$ case. We observe that at the line $q \propto N$ two pairs of PT and CC asymptotes have the same order of magnitude. Example for $N > 1$ was shown in Fig. 5. This leads to conclusion that PT results are valid under condition $q > N$, which is complementary to the CC validity condition

$q < N$. We conclude that PT and CC asymptotes form a complete set of asymptotes for image variance. Note also that, as was discussed earlier, the effective width of variance profiles is $L/ka$ for the PT case, and $L/kr_C$ for the CC case. At the PT - CC boundary $q \propto N$ both widths are of the same order, indicating that asymptotes are complete not only for the edge variance values, but uniformly for full variance profiles.

Validity domain for certain asymptotes at the $(N,q)$ plane are marked as PT1 and PT2 for the PT results for $N > 1$ and $N < 1$ correspondingly. The CC domains are marked as CC1 and CC2 for $qN > 1$ and $qN < 1$ correspondingly. The phase screen model of Section 3 has only PT1 and CC1 asymptotes. They also form a complete set of asymptotes at the $(N,q)$ plane with $q \propto N$ boundary between them. In fact the $q/N$ ratio is the single parameter for the phase screen model.

Edge variances and profile widths are summarized in Table 1 in terms of dimensional and physical parameters for $\alpha = 5/3$.

**Table 1. Summary of asymptotic results**

| Domain | Expansion type | Constraints | Edge variance | Variance profile width |
|---|---|---|---|---|
| PT1 | Perturbation theory | $1 < N < q$ | $0.266\, C_n^2 k^2 L a^{5/3}$ | $L/ka$ |
| PT2 | | $N < 1 \cap N > q$ | $0.518\, C_n^2 k^{7/3} L^{2/3} a^{7/3}$ | |
| CC1 | Coherence channels | $1/N < q < N$ | $0.565 \left(C_n^2\right)^{-1/5} k^{-2/5} L^{-1/5} a^{-1/3}$ | $L/kr_C$ |
| CC2 | | $N < q < 1/N$ | $O\!\left(\left(C_n^2\right)^{-2/5} k^{-7/15} L^{-11/15}\right)$ | |

For passive scintillometer application PT1 and PT2 domains, where image variance is directly proportional to the $C_n^2$ are the most promising, while under CC conditions the variance is less sensitive to the $C_n^2$. This limits somewhat the length of propagation paths and aperture sizes. PT1 and PT2 equations offer aperture and spectral diversions, which will be used in the measurements described in the next Section.

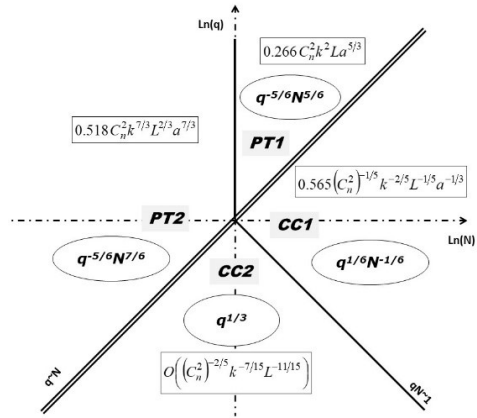

Fig. 8. Map of the image variance asymptotes. Double diagonal line $q \propto N$ represents the boundary between the PT and CC asymptotes. Solid lines – boundaries between the PT and CC subdomains. Expression in ovals represent order-of-magnitude values for Kolmogorov case in terms of dimensionless parameters $(N, q)$. Expressions in rectangles are exact asymptotes in term of physical parameters.

## 5. Measurements

Our measurements were performed on May 4, 2021 in Erie, CO around 12 PM under cloudy conditions with light wind. Target consisted of two 2'x4' sheets of plywood that were spray painted matte white and black. We used consumer-grade Canon EOS Rebel T2i camera with Opteka 650-1300 mm telephoto zoom lens. Lens was set at the maximum 950 mm focal distance. Camera was mounted on a weighted work bench at about 0.8 m above the ground, Fig. 9.

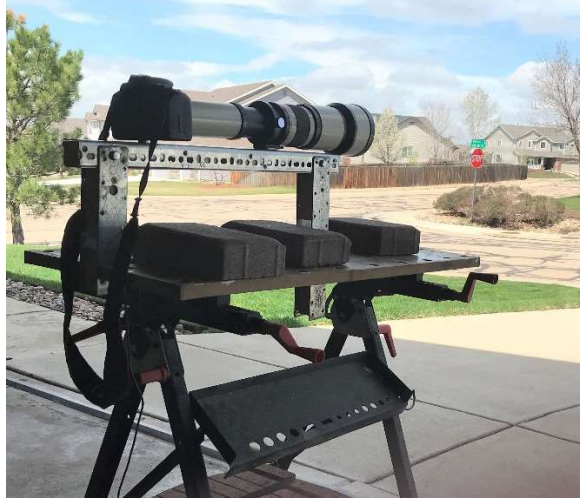

Fig. 9. Camera and lens mount.

Target images were taken at the 120 m path, mostly over the law and natural grass surface. Target was at about 1.8m above the ground, but line of sight could be as low as 0.5 m above the grass.

Camera was focused manually at the target using the largest $f/12$ aperture, and was not adjusted for the rest of the trial. Nominal aperture, as per lens specifications is about 80 mm. In order to take advantage of the aperture diversity the aperture masks with 25.4, 12.7, and 8 mm holes were placed at the entrance pupil. Exposure time and ISO#, shown in Table 2, were set manually in order to provide high recorded contrast between the black and white parts of the target..

Table 2. Camera settings [a]

| Aperture | F# | Exposure time | ISO # |
|---|---|---|---|
| 80 mm | f/12 | 1/800 | 800 |
| 25.4 mm | f/37 | 1/160 | 1600 |
| 12.7 mm | f/75 | 1/80 | 3200 |
| 8 mm | f/120 | 1/50 | 6400 |

Camera shots timing was controlled by computer. Typically, $N_F = 100$ frames at 5 seconds interval for each aperture value were taken.

Figure 10 shows on example of the original frame taken at 120 m distance. Frames were recorded as 5185x3456 pixels, .jpg files. Pixel size in the object space is about 0.52 mm.

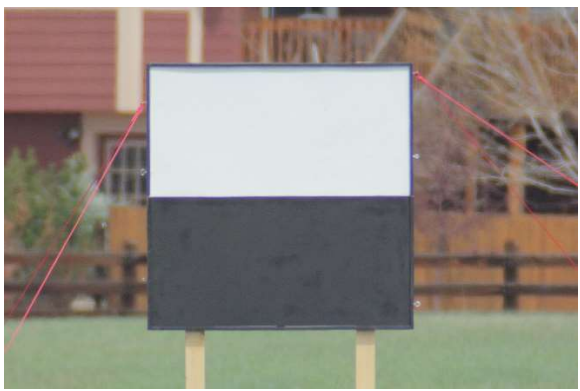

Fig. 10. Example of the original image taken by camera.

Original frames were cropped to $N_Y \times N_X$ pixels to keep only black and white parts of the target with the edge located approximately in the middle of the frame, as shown in Fig. 11. Typical $N_Y, N_X$ values are in the 2200 – 2300 range. It is worth noting that image is anisoplanatic, as evidenced by deviations of the edge from straight line.

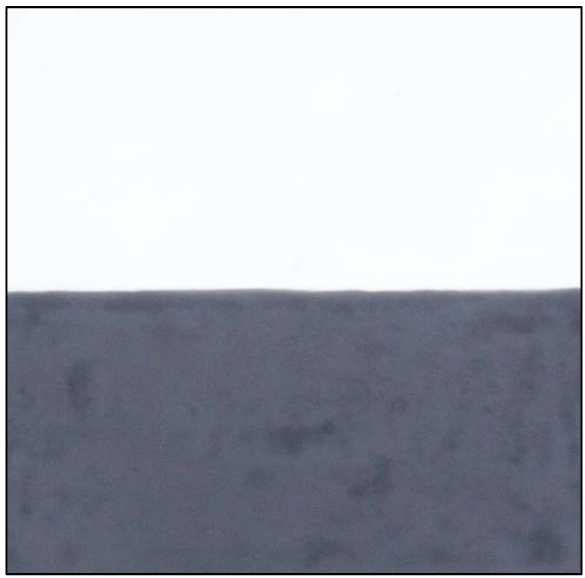

Fig.11. Example of the cropped image.

Each cropped frame is separated in three 2-D arrays

$$R_n(i,j), G_n(i,j), B_n(i,j), 1 \le i \le N_X, 1 \le j \le N_Y. \qquad (77)$$

Here $R_n, G_n, B_n$ are one byte RGB pixel readouts for *n*-th frame. The average RGB values for the dark and bright areas of each frame were calculated for each of three colors to provide parameters *B* and *C* in Eq. (4). Frame-by-frame values of these parameters can vary due to the changing illumination conditions, but we show the typical values in Table 3. The pixel readouts for the dark and bright areas varied slightly across the averaging areas with relative RMS about 3% - 5%. We believe that this was mostly due to the imperfections of the spray-painted target.

Each frame was normalized as described for Eq. (8) to bring the dark and bright areas to the nominal ±1 values.

**Table 3. Average pixel readouts and bright-dark differences for cropped frames**

|  | Red | | Green | | Blue | |
|---|---|---|---|---|---|---|
| 80 mm | 244 | 89 | 249 | 97 | 253 | 115 |
| 25.4 mm | 249 | 93 | 250 | 98 | 252 | 114 |
| 12.7 mm | 251 | 92 | 252 | 97 | 255 | 113 |
| 8 mm | 2510 | 86 | 253 | 97 | 255 | 110 |

We were not able to accurately align the target edge with the camera pixel rows with our optical setup, as evidenced from Fig. 11. The MATLAB `imrotate` function was used to achieve this alignment as follows. For several trial values of rotation angle α we calculated pixel row and frame averages and variances for the normalized images. For example for red pixels:

$$\overline{R}(i) = \frac{1}{N_F N_Y} \sum_{n=1}^{N_F} \sum_{j=1}^{N_Y} R_n(i,j), \quad \sigma_R^2(i) = \frac{1}{N_F N_Y} \sum_{n=1}^{N_F} \sum_{j=1}^{N_Y} \left[R_n(i,j) - \overline{R}(i)\right]^2. \qquad (78)$$

The optimal rotation angle was chosen based on the minimal in α value of the peak $\sigma_R^2(i)$, as illustrated in Fig. 12. For the trial described here the optimal rotation angle is $\alpha = -0.225^o$.

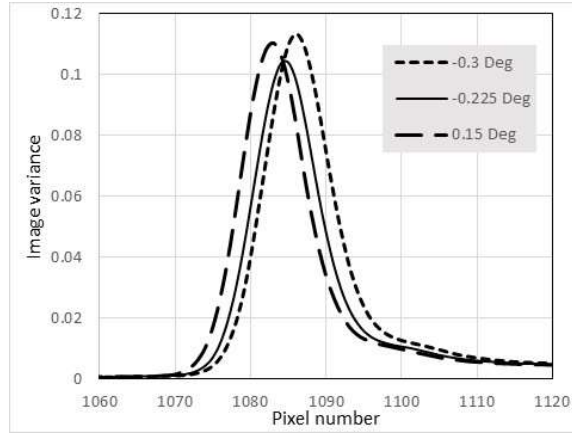

Fig.12. Example dependence of image variance on rotation angle.

Figure 13 shows vertical average profiles of the normalized RGB images, as defined by Eq. (78) for four aperture sizes. In order to compare with theoretical results, Eq. (9), data are plotted as functions of the image plane vertical coordinate $X_I$. Several observations can be made based on these data.
- Average images for full 80 mm aperture indicate the presence of chromatic aberrations.
- Average image saturates to the ±1 values away from the edge as expected.
- Asymmetry of average profiles is probably related to the inhomogeneity of the black part of target.
- The width of transition regions between the ±1 asymptotes does not exhibit expected chromatic and aperture dependencies that should be expected based on Eq. (18). This can possibly be attributed to the focusing error, or other aberrations.

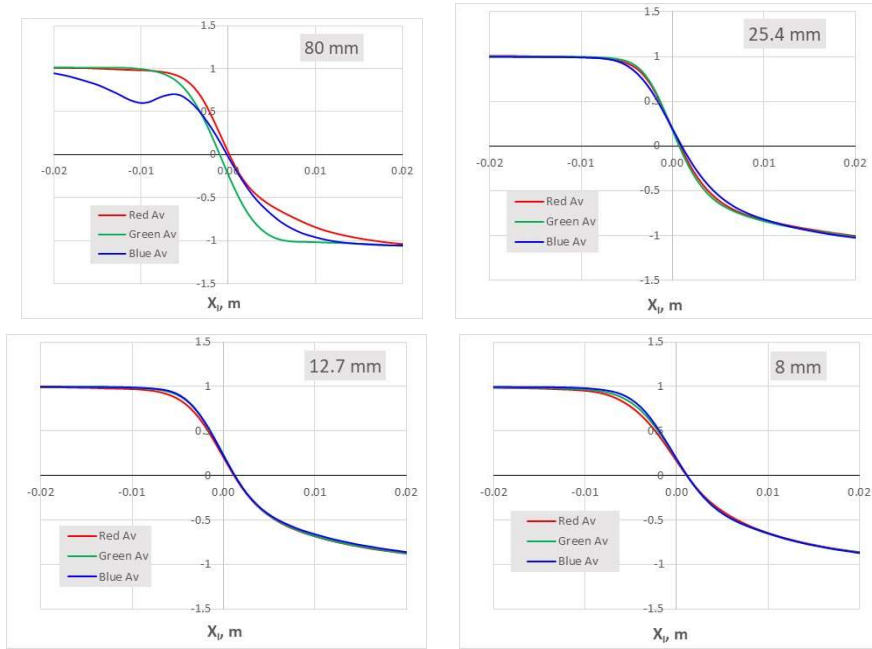

Fig.13. Average RGB images for different aperture sizes.

Figure 14 shows vertical profiles of the RGB images variances, as defined by Eq. (?) for four aperture sizes. Data are plotted as functions of the image plane vertical coordinate $X_I$. Several observations can be made.

- Image fluctuations are concentrated near the edge, as expected.
- Maximal (edge) variance does not exceed the 0.25, which is conforming to the theoretical prediction of Fig. 5.
- Chromatic dependence of the edge variance follows theoretical expectations for the smallest, 8 mm aperture only. The discrepancy is mostly caused by the blue pixels that are compromised by aberrations, as was noted earlier.

Estimations of the turbulence structure constant is based on matching theoretical and measured profile of image variance. As was elaborated in the previous sections, variance profiles depend both on the aperture size and turbulence strength. We identified four asymptotic cases for dependence of the edge variance and profile width on the turbulence strength, as shown in Fig. 1. Out of the two dimensionless parameters, $N$ and $q$ that determine fluctuations, the first one should be readily available based on the known aperture, distance and wavelength. Table 4 shows nominal values of parameter $N$. Data suggest that in most cases scintillations are in the phase screen domain.

Table 4. Nominal aperture Fresnel numbers, *N*.

|  | Red | Green | Blue |
|---|---|---|---|
| 80 mm | 126 | 162 | 185 |
| 25.4 mm | 12.4 | 15.9 | 18.1 |
| 12.7 mm | 3.1 | 4 | 4.5 |
| 8 mm | 1.2 | 1.6 | 1.8 |

However, our attempts to match the measured variance profiles to the theoretical results presented by Eq. (43) that merges the PT and CC asymptotes of the phase screen model appeared unsuccessful.

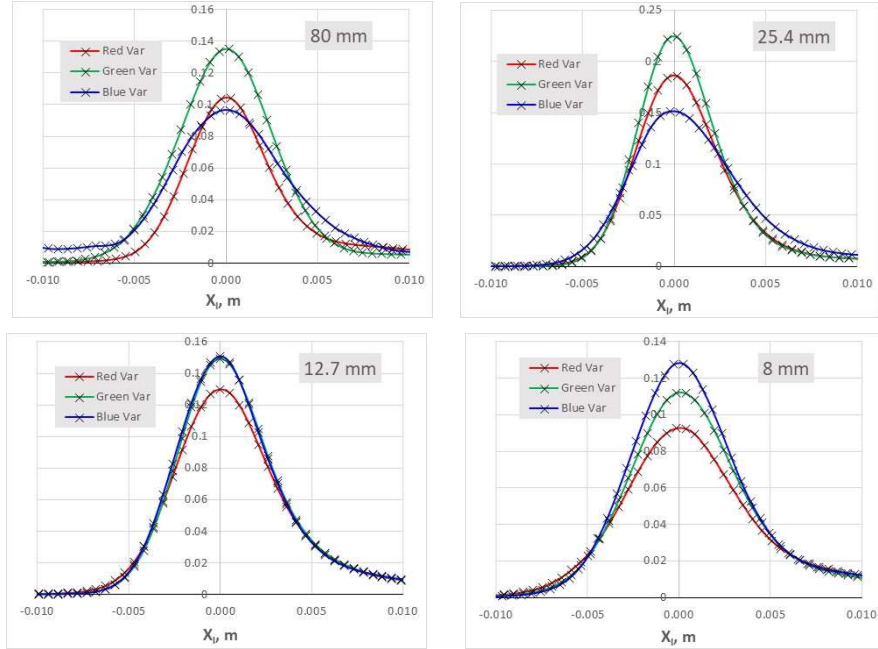

Fig.14. Measured RGB image variances for different aperture sizes.

Figure 15 shows measured variance profile for red light and 25.4 mm aperture. We matched the measured edge variance $\sigma^2(0) = 0.176$ to both PT and CC branches of the uniform approximation, Eq. (43) and Fig. 5, to arrive at two possible $a/r_C$ values of 1.2 and 36. However, the profile widths calculated based on Eq. 43 did not match the measured width, as shown in Fig. 15. The same issue occur for other colors and larger aperture sizes. We speculate that the lens aberrations and focusing error reduces the effective, in the sense of undisturbed step image width, aperture. This leads to the idea of adjusting both the effective aperture and the turbulence strength in order to match measured and theoretical profiles.

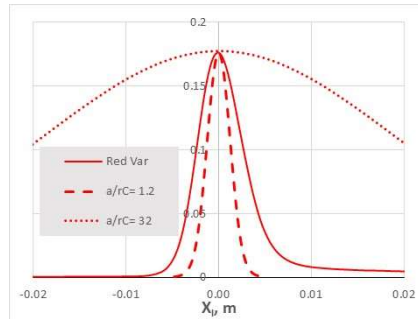

Fig.15. Measured red image variance for 25.4 mm aperture – solid curve. PT profile – long-dashed curve, CC profile – dotted curve.

Unfortunately, the aperture, and parameter $N$ are not predetermined anymore, and the location of each measurement case at the $(N, q)$ parameters plane cannot be specified beforehand. This forced us to try both uniform PS, Eq. (43) and full PT results, Eq. (60) to match the measured variance profiles. Fig. 16 and 17 show measured and theoretical PS and PT results for the largest, 80 mm and smallest 8 mm apertures. Evidently, adjustments of both the effective aperture and turbulent strength allows for a good match of the measured and theoretical profiles for both PT and PS theories.

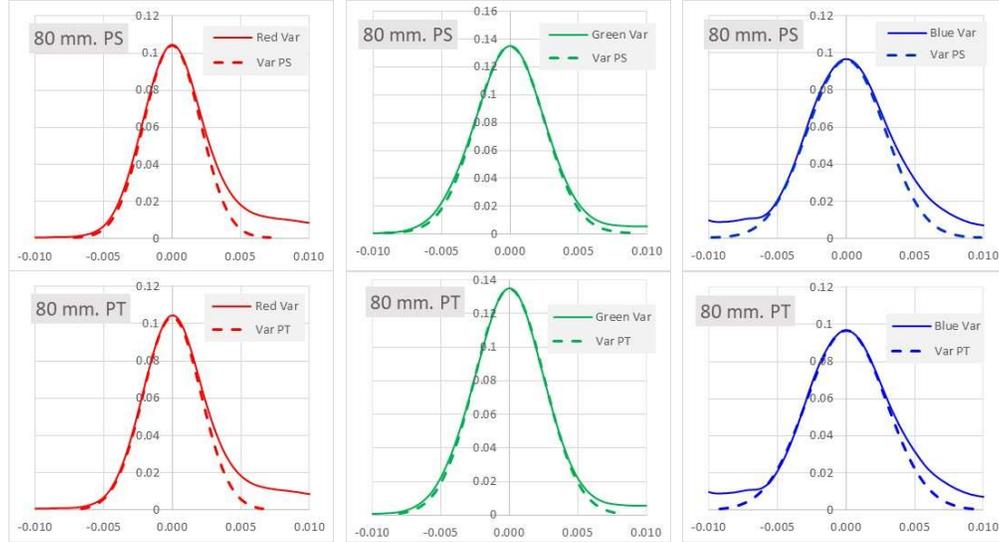

Fig.16. Matching theoretical and measured variance profiles for 80 mm aperture. Top row uniform phase screen theory, Eq. (43), bottom row perturbation theory, Eq. (60). Measured variances – solid curve, theoretical profiles – dashed curve. Argument is image plane coordinate $X_I$, m.

Fig. 18 shows effective apertures for all four nominal apertures and colors used in this trial. Dashed lines on the first two panels show the nominal aperture radii. We observe that the effective aperture is close to the nominal only for the smallest 8 mm case. For larger nominal apertures the effective sizes are almost constant, but there is a pronounced spectral dependence. The latter seems to contradict our earlier focusing error speculations. The right panel at Fig. 18 shows the scatterplot of (PS, PT) effective aperture radii. PS effective radii are consistently, but only slightly larger than for the PS theory. For smaller apertures the difference is practically negligible.

Now, when effective apertures are determined, it is possible to calculate the effective Fresnel numbers shown in Table 5, and place individual aperture and color cases on the (N, q) map. The results are shown in Fig. 19 for both PS and PT theories. Nominal asymptotic domains boundaries, $q = N$, $Nq = 1$, and $N = 1$ are shown as solid straight lines and are similar to Fig. 8. They are based on the asymptotic order-of-magnitude estimates discussed in Section 4 However these boundaries can be specified for Kolmogorov spectrum based on the exact PT1, CC1 and PT2 asymptotes given by Eq. (31), Eq. (41) and Eq. (70). We speculate that the boundary between the two asymptotic domains correspond to the equal values of corresponding asymptotes, or in other word to intersection the curves in Fig. 5 and Fig. 7.

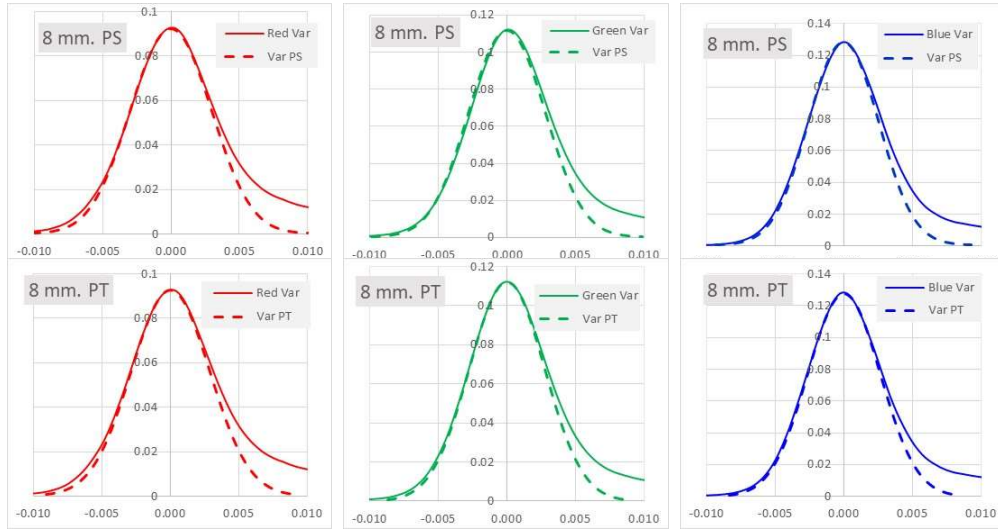

Fig.17. Matching theoretical and measured variance profiles for 8 mm aperture. Top row phase screen theory, Eq. (43), bottom row perturbation theory, Eq. (60). Measured variances – solid curve, theoretical profiles – dashed curve. Argument is image plane coordinate $X_I$, m.

Table 5. Effective aperture Fresnel numbers, *N*.

|  | PS | | | PT | | |
|---|---|---|---|---|---|---|
|  | Red | Green | Blue | Red | Green | Blue |
| 80 mm | 2.67 | 1.42 | 0.94 | 3.44 | 1.99 | 1.22 |
| 25.4 mm | 2.67 | 2.36 | 1.29 | 4.09 | 4.55 | 1.62 |
| 12.7 mm | 1.92 | 1.73 | 1.37 | 3.44 | 3.90 | 1.22 |
| 8 mm | 1.42 | 1.20 | 1.15 | 2.41 | 2.46 | 1.98 |

Thus defined boundaries are $q = 0.417N$ for the PT1 – CC1 boundary and $N = 0.413$ for the PT1 – PT2 boundary. These boundaries are shown by dashed lines in Fig. 19. After these adjustments the majority of available data points fits in the PT1 domain.

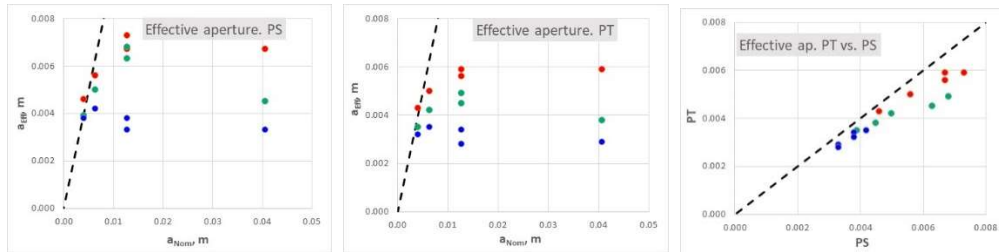

Fig.18. Left panel – dependence of the estimated effective aperture radius on nominal aperture radius for PS theory. Center panel – same for the PT theory. Right panel – relation between effective aperture radii for PS and PT theories.

Our final charts in Fig. 20 show the $C_n^2$ estimates for all colors and apertures that are based both on PS and PT theories. The larges apertures cases show wider spread of $C_n^2$ estimates, but two smaller apertures: 12.7 mm and 8 mm shown in Table 6. Here, the mean estimated value of $C_n^2 \approx 3.4 \cdot 10^{-13} \ m^{-2/3}$ is reasonable for midday observation conditions. Small chromatic variations support our color diversity expectations.

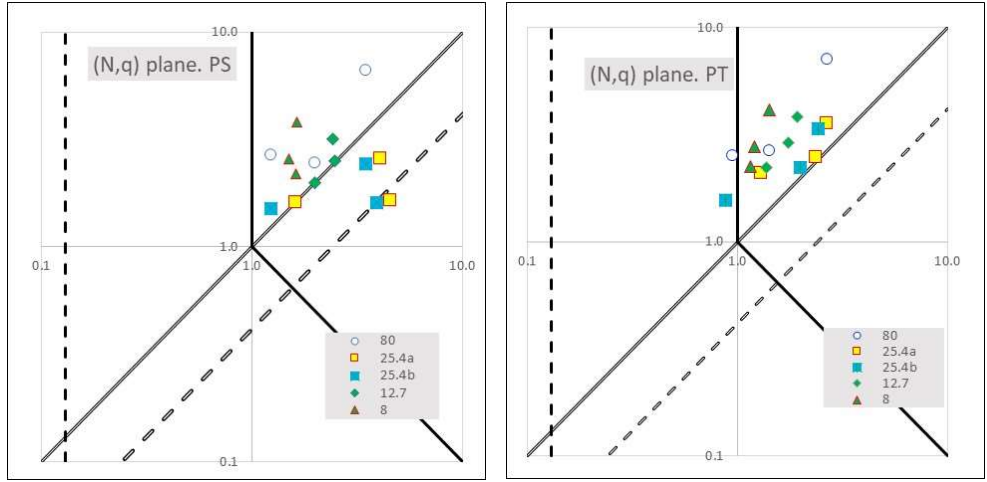

Fig.19. Left panel – dependence of the estimated effective aperture radius on nominal aperture radius for PS theory. Center panel – same for the PT theory. Right panel – relation between effective aperture radii for PS and PT theories.

## 6. Summary

We formulated the imaging problem for the incoherent step object viewed through turbulence in terms of the Green's function with distortions. The "third constraint" on the turbulent PSF allows to normalize the instantaneous images to single out the turbulence effects, and suggests that the image fluctuations are concentrated neat the edge. Statistics of the normalized images carries information about the strength of turbulence between the object and imaging aperture, and suggests its use as a passive scintillometer that does not require active light sources and can use target of opportunity in the field conditions.

A simple phase screen model was used as to introduce integration techniques and asymptotic techniques for image variance.

Complete asymptotic analysis for the extended turbulence case was based on the rigorous, path integral formulation of the Green's function. Our analysis is not restricted by isoplanatic conditions and does not use the concept of optical phase. We identified four distinct asymptotic domains for the image variance. Inadvertently, variance asymptotes in two of these domains are the same as for the phase screen model. Uniform approximation that merges these asymptotes was proposed. We identified two perturbation theory asymptotes as the most promising for the passive scintillometer application due to the linear dependence of the image variance on the turbulence structure constant.

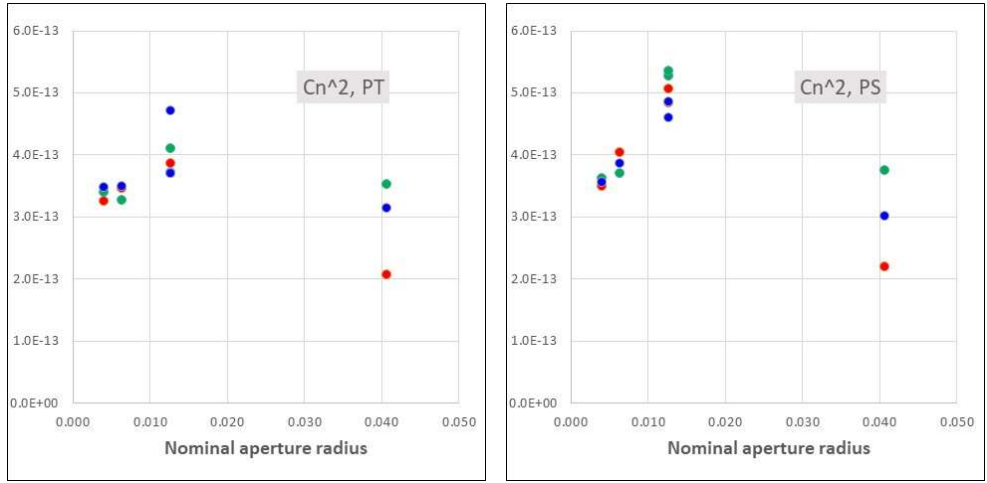

Fig.20. Estimated values of $C_n^2$, m$^{-2/3}$, for all apertures and colors. Left panel – dependence of the estimated effective aperture radius on nominal aperture radius for PS theory. Center panel – same for the PT theory. Right panel – relation between effective aperture radii for PS and PT theories.

We presented results of the field measurement that uses a shop-made target and consumer grade camera and telephoto lens. We were able to measure the image variance profiles for several apertures and three colors of camera pixels readouts. Our data qualitatively confirm theoretical expectations regarding the location and magnitude of the image fluctuations. Unfortunately, the low quality of the lens did not allow to reliably explore fluctuations for the full range of apertures. However the in the smaller aperture cases we obtained consistent across the spectrum and aperture sizes estimates of the turbulence structure constant that are reasonable for our observation conditions.

Table 6. $C_n^2$ estimates, m$^{-2/3}$, for smaller apertures

|  | PS | | | PT | | |
|---|---|---|---|---|---|---|
|  | Red | Green | Blue | Red | Green | Blue |
| 12.7 mm | 4.0E-13 | 3.7E-13 | 3.9E-13 | 3.5E-13 | 3.3E-13 | 3.5E-13 |
| 8 mm | 3.5E-13 | 3.6E-13 | 3.6E-13 | 3.2E-13 | 3.4E-13 | 3.5E-13 |

7. **Conclusions**

- Fluctuations of the normalized image of a step object can serve as a passive scintillometer, namely allow the measurement of the turbulence strength based on the fluctuation variance.
- Passive scintillometer measures turbulence strength between the object and the imaging aperture.
- Passive scintillometer does not need artificial illumination and can use specially made targets or targets of opportunity.
- Passive scintillometer generally does not require large aperture optics.
- Passive scintillometer offers aperture and chromatic diversities. They can be used to increase the accuracy of the turbulence measurements, but also can be used for estimation of the power law exponent of turbulence spectrum.

- It is possible to extend the passive scintillometer capabilities by including fluctuation statistics beyond the variance, e. g. spatial correlation of fluctuations.
- Limited proof-of-concept field trials confirm the validity of the passive scintillometer notion, but more extensive measurement under variety of turbulence conditions with higher quality optics and camera are needed.

## 8. Appendix A

Consider the object plane integral in Eq. (12)

$$\int dX_O sign(X_O) \exp\left(-\frac{ik}{L} X_O x_A\right) \tag{A1}$$

We regularize this integral by introducing a tapered step

$$sign(X_O) \Rightarrow sign(X_O) \exp\left(-\frac{|X_O|}{M}\right), \tag{A2}$$

and setting $M \to \infty$ after integral is calculated. Specifically

$$\int_{-\infty}^{\infty} dX_O sign(X_O) \exp\left(-\frac{|X_O|}{M} - \frac{ik}{L} X_O x_A\right)$$

$$= \int_0^{\infty} dX_O \left[\exp\left(-\frac{X_O}{M} - \frac{ik}{L} X_O x_A\right) - \exp\left(-\frac{X_O}{M} - \frac{ik}{L} X_O x_A\right)\right] \tag{A3}$$

$$= \frac{1}{\frac{1}{M} + \frac{ik}{L} x_A} - \frac{1}{\frac{1}{M} - \frac{ik}{L} x_A} = \frac{-2\frac{ik}{L} x_A}{\frac{1}{M^2} + \frac{k^2}{L^2} x_A^2} \underset{M \to \infty}{=} \frac{2L}{ikx_A}$$

## 9. Appendix B. Evaluation of path-integral in the PT term

First-order term of Taylor series of the last exponent in Eq. (53) for $\langle i^2(\mathbf{R}_I)\rangle$, after subtracting similar term for $\langle i(\mathbf{R}_I)\rangle^2$ is

$$\sigma_{PT}^2(\mathbf{R}_I) = \frac{k^2}{4\pi^2 L^2 \Sigma_A^2} \frac{\pi k^2}{4} \iint D^2\mathbf{v}_1(\cdot) \iint D^2\mathbf{v}_2(\cdot) \exp\left(ik\int_0^L dz \mathbf{v}_1(z)\cdot\mathbf{v}_2(z)\right) \delta\left(\int_0^L dz\mathbf{v}_1(z)\right) \delta\left(\int_0^L dz\mathbf{v}_2(z)\right)$$

$$\times \iint d^2R_{O1} sign(X_{O1}) \iint d^2R_{A1} \int d^2r_{A1} A\left(\mathbf{R}_{A1} + \frac{\mathbf{r}_{A1}}{2}\right) A\left(\mathbf{R}_{A1} - \frac{\mathbf{r}_{A1}}{2}\right) \exp\left(\frac{ik}{L}(\mathbf{R}_I - \mathbf{R}_{O1})\mathbf{r}_{A1}\right) \tag{B1}$$

$$\times \iint d^2R_{O2} sign(X_{O2}) \iint d^2R_{A2} \int d^2r_{A2} A\left(\mathbf{R}_{A2} + \frac{\mathbf{r}_{A2}}{2}\right) A\left(\mathbf{R}_{A2} - \frac{\mathbf{r}_{A2}}{2}\right) \exp\left(\frac{ik}{L}(\mathbf{R}_I - \mathbf{R}_{O2})\mathbf{r}_{A2}\right)$$

$$\int_0^L dz\left[H(\tilde{\mathbf{r}}_1(z) - \tilde{\mathbf{r}}_2(z)) + H(\tilde{\mathbf{r}}_1'(z) - \tilde{\mathbf{r}}_2'(z)) - H(\tilde{\mathbf{r}}_1(z) - \tilde{\mathbf{r}}_2'(z)) - H(\tilde{\mathbf{r}}_1'(z) - \tilde{\mathbf{r}}_2(z))\right]$$

Using spectral representation, Eq. (48) it is possible to separate the path integrals as

$$\sigma_{PT}^2(\mathbf{R}_I) = 2\pi k^2 \int_0^L dz \iint d^2\kappa \Phi_n(\kappa) \iint D^2\mathbf{v}_1(\cdot) \iint D^2\mathbf{v}_2(\cdot)$$
$$\times \exp\left(ik\int_0^L dz \mathbf{v}_1(z)\cdot\mathbf{v}_2(z) + i\kappa\int_z^L d\zeta \mathbf{v}_1(\zeta)\right) \delta\left(\int_0^L dz\mathbf{v}_1(z)\right)\delta\left(\int_0^L dz\mathbf{v}_2(z)\right) |P(\kappa,z,\mathbf{R}_I)|^2 \quad (B2)$$

Here

$$P(\kappa,z,\mathbf{R}_I) = \frac{k}{\pi L \Sigma_A} \iint d^2 R_O \operatorname{sign}(X_1) \iint d^2 R \int d^2 r A\left(\mathbf{R}+\frac{\mathbf{r}}{2}\right) A\left(\mathbf{R}-\frac{\mathbf{r}}{2}\right)$$
$$\times \exp\left[\frac{ik}{L}(\mathbf{R}_I - \mathbf{R}_O)\mathbf{r} + i\kappa\mathbf{R}_O\frac{z}{L} + i\kappa\mathbf{R}\left(1-\frac{z}{L}\right)\right] \sin\left[\frac{\kappa}{2}\left(\mathbf{r}\left(1-\frac{z}{L}\right) + \int_z^L d\zeta \mathbf{v}_2(\zeta)\right)\right] \quad (B3)$$

Integral over path variable $\mathbf{v}_1(\cdot)$ leads to the path delta-function, as detailed in the Appendix A of [9], and

$$\sigma_{PT}^2(\mathbf{R}_I) = 2\pi k^2 \int_0^L dz \iint d^2\kappa \Phi_n(\kappa) |P(\kappa,z,\mathbf{R}_I)|^2, \quad (B4)$$

where

$$P(\kappa,z,\mathbf{R}_I) = \frac{k^2}{2\pi^2 L^2 \Sigma_A} \iint d^2 R_O \operatorname{sign}(X_O) \iint d^2 R \int d^2 r A\left(\mathbf{R}+\frac{\mathbf{r}}{2}\right) A\left(\mathbf{R}-\frac{\mathbf{r}}{2}\right)$$
$$\exp\left[\frac{ik}{L}(\mathbf{R}_I - \mathbf{R}_O)\mathbf{r} + i\kappa\mathbf{R}_O\frac{z}{L} + i\kappa\mathbf{R}\left(1-\frac{z}{L}\right)\right] \sin\left[\frac{\kappa}{2}\left(\mathbf{r} - \frac{\kappa z}{k}\right)\left(1-\frac{z}{L}\right)\right] \quad (B5)$$

Integration over the object plane variable $Y_O$ simplifies Eq. (C5) to

$$P(p,q,z,X_I) = \frac{k}{\pi L \Sigma_A} \int dX_O \operatorname{sign}(X_O) \iint dX dY \int dx A\left(X+\frac{x}{2}, Y+\frac{qz}{2k}\right) A\left(X-\frac{x}{2}, Y-\frac{qz}{2k}\right)$$
$$\times \exp\left[\frac{ik}{L}(X_I - X_O)x + ipX_O\frac{z}{L} + i(pX + qY)\left(1-\frac{z}{L}\right)\right] \sin\left[\frac{1}{2}p\left(x-p\frac{z}{k}\right)\left(1-\frac{z}{L}\right)\right]. \quad (B6)$$

Integral over the object plane variable $X_O$ in Eq. (C6) can be calculated as described in Appendix A resulting in Eq. (58)

## 10. Appendix C. Transformation of Eq. (67).

After scaling spatial and spectral integration, Eq. (67) can be presented as

$$\left\langle \sigma_{PT}^2(X_I) \right\rangle_{N<1} = 2\pi C_\Phi(\alpha) C_n^2 k^2 \frac{16}{\pi^2} \int_0^L dz \left(1-\frac{z}{L}\right)^2 \iint dp dq\, p^2 (p^2+q^2)^{-1-\alpha/2}$$
$$\times \left| \int_0^\infty \frac{dt}{\left(\frac{k^2 X_I^2}{L^2}+t^2\right)} J_1^2\left(a\sqrt{\frac{k^2 X_I^2}{L^2}+t^2}\right) \cos\left(\frac{z}{k}qt\right) \right|^2 \quad (C1)$$

Integral over the spatial frequency $p$ can be calculated as follows

$$\int_{-\infty}^{\infty} dp\, p^2 (p^2+q^2)^{-1-\alpha/2} = 2q^{1-\alpha}\int_0^{\infty} dx\, x^2(1+x^2)^{-1-\alpha/2} = q^{1-\alpha}\frac{\sqrt{\pi}\,\Gamma\!\left(\dfrac{\alpha-1}{2}\right)}{2\Gamma\!\left(1+\dfrac{\alpha}{2}\right)}. \qquad (C2)$$

We use dimensionless variables $t = y/a$ and $q = x(ka/z)$ to separate physical parameters from dimensionless coefficient

$$\left\langle \sigma_{PT}^2(X_I)\right\rangle_{N<1} = C_\Phi(\alpha)\frac{32}{\sqrt{\pi}}\frac{\Gamma\!\left(\dfrac{\alpha-1}{2}\right)}{\Gamma\!\left(1+\dfrac{\alpha-1}{2}\right)}\left(C_n^2 k^2 L\right) a^{4-\alpha} k^{2-\alpha} L^{\alpha-2} \int_0^1 d\zeta\,(1-\zeta)^2 \zeta^{\alpha-2}$$

$$\times \int_0^{\infty} dx\, x^{1-\alpha}\left|\int_0^{\infty}\frac{dy}{\left(\dfrac{k^2 a^2}{L^2}X_I^2+y^2\right)} J_1^2\!\left(\sqrt{\dfrac{k^2 a^2}{L^2}X_I^2+y^2}\right)\cos(xy)\right|^2 \qquad (C3)$$

Integral over $\zeta$ is calculated as

$$\int_0^1 d\zeta\,(1-\zeta)^2 \zeta^{\alpha-2} = \frac{2}{\alpha(\alpha^2-1)}, \qquad (C4)$$

and integral over $x$ in Eq. (B3) is

$$\times \int_0^{\infty} dx\, x^{1-\alpha}\cos(xy)\cos(xy') = -\frac{1}{2}\Gamma(2-\alpha)\cos\!\left(\frac{\pi\alpha}{2}\right)\!\left(|y+y'|^{\alpha-2}+|y-y'|^{\alpha-2}\right) \qquad (C5)$$

Finally, compact form for image variance is presented as computation capable double integral

$$\left\langle \sigma_{PT}^2(X_I)\right\rangle_{N<1} = C_\Phi(\alpha)\frac{32}{\sqrt{\pi}}\frac{\Gamma\!\left(\dfrac{\alpha-1}{2}\right)}{\Gamma\!\left(1+\dfrac{\alpha}{2}\right)}\frac{\Gamma(2-\alpha)}{\alpha(\alpha^2-1)}\cos\!\left(\frac{\pi(\alpha-1)}{2}\right)\!\left(C_n^2 k^2 L\right) a^{4-\alpha} k^{2-\alpha} L^{\alpha-2}$$

$$\times \int_0^{\infty}\frac{dy}{\left(\dfrac{k^2 a^2}{L^2}X_I^2+y^2\right)} J_1^2\!\left(\sqrt{\dfrac{k^2 a^2}{L^2}X_I^2+y^2}\right)\int_0^{\infty}\frac{dy'}{\left(\dfrac{k^2 a^2}{L^2}X_I^2+y'^2\right)} J_1^2\!\left(\sqrt{\dfrac{k^2 a^2}{L^2}X_I^2+y'^2}\right) \qquad (C6)$$

$$\times \left(|y+y'|^{\alpha-2}+|y-y'|^{\alpha-2}\right)$$

## 11. Appendix D. Evaluation of path-integral in the $M_1$ term.

$M_1$ term of the CC series has the following form

$$\sigma_{M1}^2(\mathbf{R}_I) = \frac{k^2}{4\pi^2 L^2 \Sigma_A} \frac{\pi k^2}{4} \iint D^2 \mathbf{v}_1(\cdot) \iint D^2 \mathbf{v}_2(\cdot) \exp\left(ik \int_0^L dz \mathbf{v}_1(z) \cdot \mathbf{v}_2(z)\right) \delta\left(\int_0^L dz \mathbf{v}_1(z)\right) \delta\left(\int_0^L dz \mathbf{v}_2(z)\right)$$

$$\times \iint d^2 R_{O1} \text{sign}(X_{O1}) \iint d^2 R_{A1} \int d^2 r_{A1} A\left(\mathbf{R}_{A1} + \frac{\mathbf{r}_{A1}}{2}\right) A\left(\mathbf{R}_{A1} - \frac{\mathbf{r}_{A1}}{2}\right) \exp\left(\frac{ik}{L}(\mathbf{R}_I - \mathbf{R}_{O1})\mathbf{r}_{A1}\right)$$

$$\times \iint d^2 R_{O2} \text{sign}(X_{O2}) \iint d^2 R_{A2} \int d^2 r_{A2} A\left(\mathbf{R}_{A2} + \frac{\mathbf{r}_{A2}}{2}\right) A\left(\mathbf{R}_{A2} - \frac{\mathbf{r}_{A2}}{2}\right) \exp\left(\frac{ik}{L}(\mathbf{R}_I - \mathbf{R}_{O2})\mathbf{r}_{A2}\right) \quad \text{(D1)}$$

$$\times \exp\left\{-\frac{\pi k^2}{4} \int_0^L dz \left[H(\tilde{\mathbf{r}}_1(z) - \tilde{\mathbf{r}}_1'(z)) + H(\tilde{\mathbf{r}}_2(z) - \tilde{\mathbf{r}}_2'(z))\right]\right\}$$

$$\times \int_0^L dz \left[H(\tilde{\mathbf{r}}_1(z) - \tilde{\mathbf{r}}_2(z)) + H(\tilde{\mathbf{r}}_1'(z) - \tilde{\mathbf{r}}_2'(z)) - H(\tilde{\mathbf{r}}_1(z) - \tilde{\mathbf{r}}_2'(z)) - H(\tilde{\mathbf{r}}_1'(z) - \tilde{\mathbf{r}}_2(z))\right]$$

Using spectral representation, Eq. (52) it is possible to separate the path integrals as

$$\sigma_{M1}^2(\mathbf{R}_I) = 2\pi k^2 \int_0^L dz \iint d^2\kappa \Phi_n(\boldsymbol{\kappa}) \iint D^2 \mathbf{v}_1(\cdot) \iint D^2 \mathbf{v}_2(\cdot)$$

$$\times \exp\left(ik \int_0^L dz \mathbf{v}_1(z) \cdot \mathbf{v}_2(z) + i\boldsymbol{\kappa} \int_z^L d\zeta \mathbf{v}_1(\zeta)\right) \delta\left(\int_0^L dz \mathbf{v}_1(z)\right) \delta\left(\int_0^L dz \mathbf{v}_2(z)\right) |M_1(\boldsymbol{\kappa}, z, \mathbf{R}_I)|^2 \quad \text{(D2)}$$

Here

$$M_1(\boldsymbol{\kappa}, z, \mathbf{R}_I) = \frac{k}{\pi L \Sigma_A} \iint d^2 R_O \text{sign}(X_1) \iint d^2 R_A \int d^2 r_A A\left(\mathbf{R}_A + \frac{\mathbf{r}_A}{2}\right) A\left(\mathbf{R}_A - \frac{\mathbf{r}_A}{2}\right)$$

$$\exp\left[\frac{ik}{L}(\mathbf{R}_I - \mathbf{R}_O)\mathbf{r}_A + i\boldsymbol{\kappa}\mathbf{R}_O \frac{z}{L} + i\boldsymbol{\kappa}\mathbf{R}_A\left(1 - \frac{z}{L}\right)\right] \sin\left[\frac{\boldsymbol{\kappa}}{2}\left(\mathbf{r}_A\left(1 - \frac{z}{L}\right) + \int_z^L d\zeta \mathbf{v}_2(\zeta)\right)\right] \quad \text{(D3)}$$

$$\exp\left[-\frac{\pi k^4}{4} \int_0^L dz' H\left(\mathbf{r}_A\left(1 - \frac{z'}{L}\right) + \int_{z,}^L d\zeta \mathbf{v}_2(\zeta)\right)\right]$$

Integral over path variable $\mathbf{v}_1(\cdot)$ leads to the path delta-function, as detailed in the Appendix A of [9], and

$$\sigma_{M1}^2(\mathbf{R}_I) = 2\pi k^2 \int_0^L dz \iint d^2\kappa \Phi_n(\boldsymbol{\kappa}) |M_2(\boldsymbol{\kappa}, z, \mathbf{R}_I)|^2, \quad \text{(D4)}$$

where

$$M_2(\boldsymbol{\kappa}, z, \mathbf{R}_I) = \frac{k^2}{2\pi^2 L^2 \Sigma_A} \iint d^2 R_O \text{sign}(X_O) \iint d^2 R \int d^2 r A\left(\mathbf{R} + \frac{\mathbf{r}}{2}\right) A\left(\mathbf{R} - \frac{\mathbf{r}}{2}\right)$$

$$\times \exp\left[\frac{ik}{L}(\mathbf{R}_I - \mathbf{R}_O)\mathbf{r} + i\boldsymbol{\kappa}\mathbf{R}_O \frac{z}{L} + i\boldsymbol{\kappa}\mathbf{R}\left(1 - \frac{z}{L}\right)\right] \sin\left[\frac{\boldsymbol{\kappa}}{2}\left(\mathbf{r} - \frac{\boldsymbol{\kappa} z}{k}\right)\left(1 - \frac{z}{L}\right)\right] \quad \text{(D5)}$$

$$\times \exp\left[-\frac{\pi k^4}{4} \int_0^L dz' H\left(\mathbf{r}_A\left(1 - \frac{z'}{L}\right) - \boldsymbol{\kappa} \frac{\min(z, z')}{L}\left(1 - \frac{\max(z, z')}{L}\right)\right)\right]$$

Integration over the object plane variable $Y_O$ simplifies Eq. (D5) to

$$M_2(p,q,z,X_I) = \frac{k}{\pi L \Sigma_A} \int dX_O \operatorname{sign}(X_O) \iint dXdY \int dx A\left(X + \frac{x}{2} + \frac{pz}{2k}, Y_A + \frac{qz}{2k}\right)$$
$$\times A\left(X - \frac{x}{2} - \frac{pz}{2k}, Y_A - \frac{qz}{2k}\right) \sin\left[\frac{1}{2} px\left(1 - \frac{z}{L}\right)\right] \exp\left[-\frac{\pi k^2}{4} \int_0^L d\zeta H[\mathbf{r}(z,\zeta)]\right] \quad \text{(D6)}$$
$$\times \exp\left[\frac{ik}{L} X_I\left(x + p\frac{z}{k}\right) - \frac{ik}{L} X_O x + ipX_A\left(1 - \frac{z}{L}\right) + iqY_A\left(1 - \frac{z}{L}\right)\right],$$

where

$$\mathbf{r}(z,\zeta) \equiv \left(x\left(1 - \frac{\zeta}{L}\right) + \frac{p}{k}[z - \min(z,\zeta)], \frac{q}{k}[z - \min(z,\zeta)]\right) \quad \text{(D7)}$$

Integral over the object plane variable $X_O$ in Eq. (D6) can be calculated as described in Appendix A resulting in Eq. (71).

## 12. Appendix E. Evaluation of path integral in the $A_0$ term.

We use analogy with Eq. (31) in Eq. (53) to write path-integral for the $A_0$ term as

$$\langle i^2(\mathbf{R}_I)\rangle_{A0} = \frac{k^2}{4\pi^2 L^2 \Sigma_A^2} \iint D^2\mathbf{v}_1(\cdot) \iint D^2\mathbf{v}_2(\cdot) \exp\left(ik \int_0^L dz \mathbf{v}_1(z) \cdot \mathbf{v}_2(z)\right) \delta\left(\int_0^L dz \mathbf{v}_1(z)\right) \delta\left(\int_0^L dz \mathbf{v}_2(z)\right)$$

$$\times \iint d^2 R_{O1} \operatorname{sign}(X_{O1}) \iint d^2 R_{A1} \int d^2 r_{A1} A\left(\mathbf{R}_{A1} + \frac{\mathbf{r}_{A1}}{2}\right) A\left(\mathbf{R}_{A1} - \frac{\mathbf{r}_{A1}}{2}\right) \exp\left(\frac{ik}{L}(\mathbf{R}_I - \mathbf{R}_{O1})\mathbf{r}_{A1}\right)$$

$$\times \iint d^2 R_{O2} \operatorname{sign}(X_{O2}) \iint d^2 R_{A2} \int d^2 r_{A2} A\left(\mathbf{R}_{A2} + \frac{\mathbf{r}_{A2}}{2}\right) A\left(\mathbf{R}_{A2} - \frac{\mathbf{r}_{A2}}{2}\right) \exp\left(\frac{ik}{L}(\mathbf{R}_I - \mathbf{R}_{O2})\mathbf{r}_{A2}\right) \quad \text{(E1)}$$

$$\times \exp\left\{-\frac{\pi k^2}{4} \int_0^L dz \left[H\left((\mathbf{R}_{O1} - \mathbf{R}_{O2})\frac{z}{L} + \left(\mathbf{R}_{A1} - \mathbf{R}_{A2} + \frac{\mathbf{r}_{A2} + \mathbf{r}_{A2}}{2}\right)\left(1 - \frac{z}{L}\right) + \int_z^L d\zeta \mathbf{v}_1(\zeta)\right) \right. \right.$$
$$\left. \left. + H\left((\mathbf{R}_{O1} - \mathbf{R}_{O2})\frac{z}{L} + \left(\mathbf{R}_{A1} - \mathbf{R}_{A2} - \frac{\mathbf{r}_{A2} + \mathbf{r}_{A2}}{2}\right)\left(1 - \frac{z}{L}\right) + \int_z^L d\zeta \mathbf{v}_1(\zeta)\right)\right]\right\}.$$

Integral over the $\mathbf{v}_2(\cdot)$ leads to the path $\delta$ – function, Eq. ( ), and after a simple change of the image and aperture plane variables Eq. (E1) reduces to

$$\langle i^2(\mathbf{R}_I)\rangle_{A0} = \frac{k^4}{16\pi^4 L^4 \Sigma_A^2} \iint d^2 R_O \iint d^2 r_O \operatorname{sign}\left(X_O + \frac{x_0}{2}\right) \operatorname{sign}\left(X_O - \frac{x_0}{2}\right)$$

$$\times \iint d^2 R \iint d^2 r_1 \iint d^2 r_2 \iint d^2 \rho A\left(\mathbf{R} + \frac{\mathbf{r}_1}{2} + \frac{\mathbf{r}_2}{2} + \frac{\boldsymbol{\rho}}{4}\right) A\left(\mathbf{R} + \frac{\mathbf{r}_1}{2} - \frac{\mathbf{r}_2}{2} - \frac{\boldsymbol{\rho}}{4}\right)$$

$$\times A\left(\mathbf{R} - \frac{\mathbf{r}_1}{2} + \frac{\mathbf{r}_2}{2} - \frac{\boldsymbol{\rho}}{4}\right) A\left(\mathbf{R} + \frac{\mathbf{r}_1}{2} - \frac{\mathbf{r}_2}{2} + \frac{\boldsymbol{\rho}}{4}\right) \exp\left(\frac{ik}{L}(\mathbf{R}_I - \mathbf{R}_O)\boldsymbol{\rho} - \frac{ik}{L}\mathbf{r}_2\mathbf{r}_O\right) \quad \text{(E2)}$$

$$\times \exp\left\{-\frac{\pi k^2}{4} \int_0^L dz \left[H\left(\mathbf{r}_O \frac{z}{L} + \left(\mathbf{r}_1 + \frac{\boldsymbol{\rho}}{2}\right)\left(1 - \frac{z}{L}\right)\right) + H\left(\mathbf{r}_O \frac{z}{L} + \left(\mathbf{r}_1 - \frac{\boldsymbol{\rho}}{2}\right)\left(1 - \frac{z}{L}\right)\right)\right]\right\}.$$

As was already stated when evaluating the $M_1$ term, CC expansion requires that $r_C < a$, and

$$\langle i^2(\mathbf{R}_I)\rangle_{A0} \approx \frac{k^4}{16\pi^4 L^4 \Sigma_A^2} \iint d^2R_O \iint d^2r_O \, \text{sign}\left(X_O + \frac{x_0}{2}\right)\text{sign}\left(X_O - \frac{x_0}{2}\right)$$

$$\times \iint d^2R \iint d^2r_1 \iint d^2r_2 \iint d^2\rho \, A^2\left(\mathbf{R} + \frac{\mathbf{r}_2}{2}\right)A^2\left(\mathbf{R} - \frac{\mathbf{r}_2}{2}\right)\exp\left(\frac{ik}{L}(\mathbf{R}_I - \mathbf{R}_O)\boldsymbol{\rho} - \frac{ik}{L}\mathbf{r}_2\mathbf{r}_O\right) \quad (E3)$$

$$\times \exp\left\{-\frac{\pi k^2}{4}\int_0^L dz\left[H\left(\mathbf{r}_O\frac{z}{L} + \left(\mathbf{r}_1 + \frac{\boldsymbol{\rho}}{2}\right)\left(1 - \frac{z}{L}\right)\right) + H\left(\mathbf{r}_O\frac{z}{L} + \left(\mathbf{r}_1 - \frac{\boldsymbol{\rho}}{2}\right)\left(1 - \frac{z}{L}\right)\right)\right]\right\}.$$

After integration over $Y_0$, Eq. (E3) can be presented as

$$\langle i^2(X_I)\rangle_{A0} \approx \frac{k^3}{8\pi^3 L^3 \Sigma_A^2}\iint dX_O dY_O \iint dx_O dy_O \, \text{sign}\left(X_O + \frac{x_0}{2}\right)\text{sign}\left(X_O - \frac{x_0}{2}\right)$$

$$\times \iint dXdY \iint dx_2 dy_2 \int d\xi \, A^2\left(X + \frac{x_2}{2}, Y + \frac{y_2}{2}\right)A^2\left(X - \frac{x_2}{2}, Y - \frac{y_2}{2}\right)$$

$$\exp\left[\frac{ik}{L}(X_I\xi - X_O\xi) - \frac{ik}{L}(x_2 x_O + y_2 y_O)\right] \quad (E4)$$

$$\times \iint dx_1 dy_1 \exp\left\{-\frac{\pi k^2}{4}\int_0^L dz \begin{bmatrix} H\left(x_O\frac{z}{L} + \left(x_1 + \frac{\xi}{2}\right)\left(1 - \frac{z}{L}\right), y_O\frac{z}{L} + y_1\left(1 - \frac{z}{L}\right)\right) \\ + H\left(x_O\frac{z}{L} + \left(x_1 - \frac{\xi}{2}\right)\left(1 - \frac{z}{L}\right), y_O\frac{z}{L} + y_1\left(1 - \frac{z}{L}\right)\right) \end{bmatrix}\right\}.$$

Eq. (E4) indicates that, just as the $M_1$ term, effective width of the $\langle i^2(X_I)\rangle_{A0}$ profile is $O(L/kr_C)$, and we limit further development to the edge location, $X_I = 0$. Two asymptotic cases can be identified.

When $r_C > L/ka$, variables $x_O, y_O \propto L/ka < r_C$ and Eq. (D4) is simplified to

$$\langle i^2(0)\rangle_{A0} \approx \frac{k^2}{4\pi^2 L^2 \Sigma_A^2}\int dX_O \int dx_O \, \text{sign}\left(X_O + \frac{x_0}{2}\right)\text{sign}\left(X_O - \frac{x_0}{2}\right)$$

$$\times \iint dXdY \int dx_2 \int d\xi \, A^2\left(X + \frac{x_2}{2}, Y\right)A^2\left(X - \frac{x_2}{2}, Y\right)\exp\left[-\frac{ik}{L}(X_O\xi + x_2 x_O)\right]$$

$$\times \iint dx_1 dy_1 \exp\left\{-\frac{\pi k^2}{4}\int_0^L dz\left[H\left(\left(x_1 + \frac{\xi}{2}\right)\left(1 - \frac{z}{L}\right), y_1\left(1 - \frac{z}{L}\right)\right) + H\left(\left(x_1 - \frac{\xi}{2}\right)\left(1 - \frac{z}{L}\right), y_1\left(1 - \frac{z}{L}\right)\right)\right]\right\}.$$

(E5)

Variables $X, Y, x_2$ in Eq. (E5) are of order of the aperture size $a$, and $X_O \propto L/kr_C$, and integral is estimated as

$$\langle i^2(0)\rangle_{A0} = O\left(\frac{q}{N}\right) \quad (E6)$$

When $r_C < L/ka$, variables $x_O, y_O \propto r_C < L/ka$ and Eq. (E4) is simplified to

$$\langle i^2(0)\rangle_{A0} \approx \frac{k^3}{8\pi^3 L^3}\iint dX_O dx_O \int d\xi\, sign\left(X_O+\frac{x_0}{2}\right)sign\left(X_O-\frac{x_0}{2}\right)\exp\left(-\frac{ik}{L}X_O\xi\right)$$

$$\times \iint dx_1 dy_1 \int dy_O \exp\left\{-\frac{\pi k^2}{4}\int_0^L dz\left[\begin{array}{l}H\left(x_O\frac{z}{L}+\left(x_1+\frac{\xi}{2}\right)\left(1-\frac{z}{L}\right), y_O\frac{z}{L}+y_1\left(1-\frac{z}{L}\right)\right)\\+H\left(x_O\frac{z}{L}+\left(x_1-\frac{\xi}{2}\right)\left(1-\frac{z}{L}\right), y_O\frac{z}{L}+y_1\left(1-\frac{z}{L}\right)\right)\end{array}\right]\right\}.$$
(E7)

Variables $x_1, y_1$ in Eq. (E7) are of order of coherence radius $r_C$, and $X_O \propto L/kr_C$, and integral is estimated as

$$\langle i^2(0)\rangle_{A0} = O(q^2)$$
(E8)